\documentclass[lettersize,journal]{IEEEtran}

\title{\textsc{REPOSE}: Quantifying the Price of Security in Weakly-Hard Real-Time Cyber-Physical Systems}

\usepackage{amsmath,amsfonts}
\usepackage{graphicx}
\usepackage{textcomp}
\usepackage{comment}
\usepackage{algorithm}
\usepackage{xcolor}

\usepackage{flushend}
\usepackage[noend]{algpseudocode}
\usepackage{array}
\usepackage{url}
\usepackage{listings}
\usepackage{verbatim}
\usepackage{soul}
\usepackage{xspace}
\usepackage{subcaption}
\usepackage[colorlinks=true,
            citecolor=blue,
            linkcolor=black,
            urlcolor=black]{hyperref}

\usepackage{orcidlink}

\author{
    Vijay Banerjee\,\orcidlink{0000-0001-5562-8469} and
    Monowar Hasan\,\orcidlink{0000-0002-2657-0402}%
   \thanks{This work was supported by the National Science Foundation (NSF) under Grant 2442595.\textit{(Corresponding authors: V. Banerjee; M. Hasan)}}
    \thanks{Vijay Banerjee and Monowar Hasan are with the School of Electrical Engineering and Computer Science, Washington State University, Pullman, WA, USA (emails: vijay.banerjee@wsu.edu; monowar.hasan@wsu.edu).}
    
}

\begin{document}

\maketitle

\newcommand{\da}{\downarrow}
\newcommand{\la}{\leftarrow}
\newcommand{\ra}{\rightarrow}
\newcommand{\lla}{\longleftarrow}
\newcommand{\llra}{\longleftrightarrow}
\newcommand{\lra}{\longrightarrow}
\newcommand{\Lla}{\Longleftarrow}
\newcommand{\Llra}{\Longleftrightarrow}
\newcommand{\Lra}{\Longrightarrow}
\newcommand{\ua}{\uparrow}

\newcommand{\nea}{\nearrow}
\newcommand{\nwa}{\nwarrow}
\newcommand{\sea}{\searrow}
\newcommand{\swa}{\swarrow}

\newcommand{\mf}{\mathbf}
\newcommand{\mc}{\mathcal}
\newcommand{\ol}{\overline}

\newcommand{\eg}{{e.g.,}\xspace}
\newcommand{\viz}{{viz.,}\xspace}
\newcommand{\Eg}{{E.g.,}\xspace}
\newcommand{\etal}{{et~al.}\xspace}
\newcommand{\ie}{{i.e.,}\xspace}
\newcommand{\etc}{{etc.}\xspace}
\newcommand{\ci}{{\it (i) }}
\newcommand{\cii}{{\it (ii) }}
\newcommand{\ciii}{{\it (iii) }}
\newcommand{\civ}{{\it (iv) }}
\newcommand{\cv}{{\it (v) }}
\newcommand{\cvi}{{\it (vi) }}
\newcommand{\cvii}{{\it (vii) }}
\newcommand{\ca}{{\it (a) }}
\newcommand{\cb}{{\it (b) }}
\newcommand{\cc}{{\it (c) }}
\newcommand{\cd}{{\it (d) }}
\newcommand{\ce}{{\it (e) }}
\newcommand{\cf}{{\it (f) }}
\newcommand{\cg}{{\it (g) }}
\newcommand{\ch}{{\it (h) }}
\newcommand{\cj}{{\it (j) }}
\newcommand{\ck}{{\it (k) }}
\newcommand{\cl}{{\it (l) }}
\newcommand{\cm}{{\it (m) }}
\newcommand{\cn}{{\it (n) }}
\newcommand{\co}{{\it (o) }}
\newcommand{\cp}{{\it (p) }}
\newcommand{\cq}{{\it (q) }}
\newcommand{\mnote}{{\ul{Note:}}\xspace}

\newcommand{\cpar}[1]{{\vspace*{0.5\baselineskip}\noindent\bfseries #1\;}}


\newcolumntype{L}[1]{>{\raggedright\let\newline\\\arraybackslash\hspace{0pt}}m{#1}}
\newcolumntype{C}[1]{>{\centering\let\newline\\\arraybackslash\hspace{0pt}}m{#1}}
\newcolumntype{R}[1]{>{\raggedleft\let\newline\\\arraybackslash\hspace{0pt}}m{#1}}
\newcolumntype{P}[1]{>{\raggedright\arraybackslash}p{#1}}

\definecolor{notecolor}{rgb}{0.8,0,0} 
\newcommand\mhnote[1]{{\textcolor{notecolor}{[\textit{#1} --- \textbf{MH}]}}}

\newcommand{\repose}{\textsc{REPOSE}}
\newcommand{\restart}{\textsc{Cleanup+Restart}}
\newcommand{\cleanup}{\textsc{Cleanup}}
\newcommand{\onlyrestart}{\textsc{Restart}}
\newtheorem{theorem}{Theorem}
\newtheorem{lemma}{Lemma}
\newtheorem{definition}{Definition}

\begin{abstract}
In contemporary IoT edge devices with real-time requirements, security is primarily enforced through design-time parameters associated with security tasks, leading to mechanisms that operate in an \emph{opportunistic} manner. As a result, security checks are often performed as secondary operations. This approach can result in systems where no security tasks are executed due to high utilization by other tasks. An alternative approach taken in prior work is to add security mechanisms to every task in the system, resulting in substantially lower performance than that of a system with no security. These approaches have resulted in an \emph{all-or-nothing} scenario for edge device security, motivating numerous studies on the safety-security trade-off in real-time cyber-physical systems (RT-CPS). This study introduces an analytical framework---\repose---for evaluating the security feasibility of real-time control systems at runtime.  \repose\ is developed for \textit{weakly-hard} real-time control systems that facilitate a ``bounded trade-off'' between safety and security. In contrast to imposing additional (pessimistic) design-time overhead as considered in some real-time security literature, \repose\ performs security operations in both \textit{proactive} and \textit{reactive} manners based on the task's current behavior. Our evaluations show that \repose\ can effectively add security operations to RT-CPS with a feasibility overhead of $0.06\%$ at $80\%$ utilization, compared to a $ 29\%$ overhead observed in systems with hard constraints. Through a case study of a classic control system, we also demonstrate that \repose\ provides a robust framework to \textit{analyze and calculate} the safety-security tradeoff.
\end{abstract}


\section{Introduction}
\IEEEPARstart{S}{ecurity} integration in most real-time cyber-physical systems (RT-CPS) assumes a system model with \emph{hard} constraints~\cite{hasan2024sok}. However, many RT-CPS are capable of handling occasional (but \textit{bounded}) deadline misses without compromising the system's safety~\cite{maggio2020control,pazzaglia_dmac_2019,vreman2021stability}. RT-CPS with bounded deadline misses are known as weakly-hard systems~\cite{bernat2001weakly}. Prior studies have shown that security mechanisms designed for ``hard'' real-time systems are overly restrictive and may make security integration infeasible, especially in highly utilized systems~\cite{lin2015security}. Despite a decade of research on scheduler-level techniques for RT-CPS security~\cite{hasan2024sok}, the integration of security into weakly-hard systems remains largely unexplored.

In this paper, we introduce \repose, a framework that formally quantifies the ``price'' of security mechanisms when integrating them into RT-CPS.\footnote{\repose\ naming is originated from the phrase {\em \textbf{\ul{re}}al-time systems' \textbf{\ul{p}}rice \textbf{\ul{o}}f \textbf{\ul{se}}curity}.} We built \repose\ for RT-CPS with weakly-hard requirements,  often characterized as $(m,k)$-constraints, where a task can tolerate up to $m$ missed deadlines in every $k$ invocations without entering an unsafe state~\cite{bernat2001weakly}. \repose\ allows us to \textit{quantify} the security overhead a system can \textit{safely} accommodate while still adhering to its weakly-hard constraints. Notably, in many IoT environments, security tasks do not necessarily require strict periodic invocation, unlike other real-time or control tasks. Hence, during a security event, occasional skips can be tolerated as long as they remain within the task's weakly-hard requirements (e.g., follow $(m,k)$ semantics). This flexibility enables better admissibility for security monitoring tasks.

\begin{figure*}
    \centering
    \includegraphics[width=\textwidth]{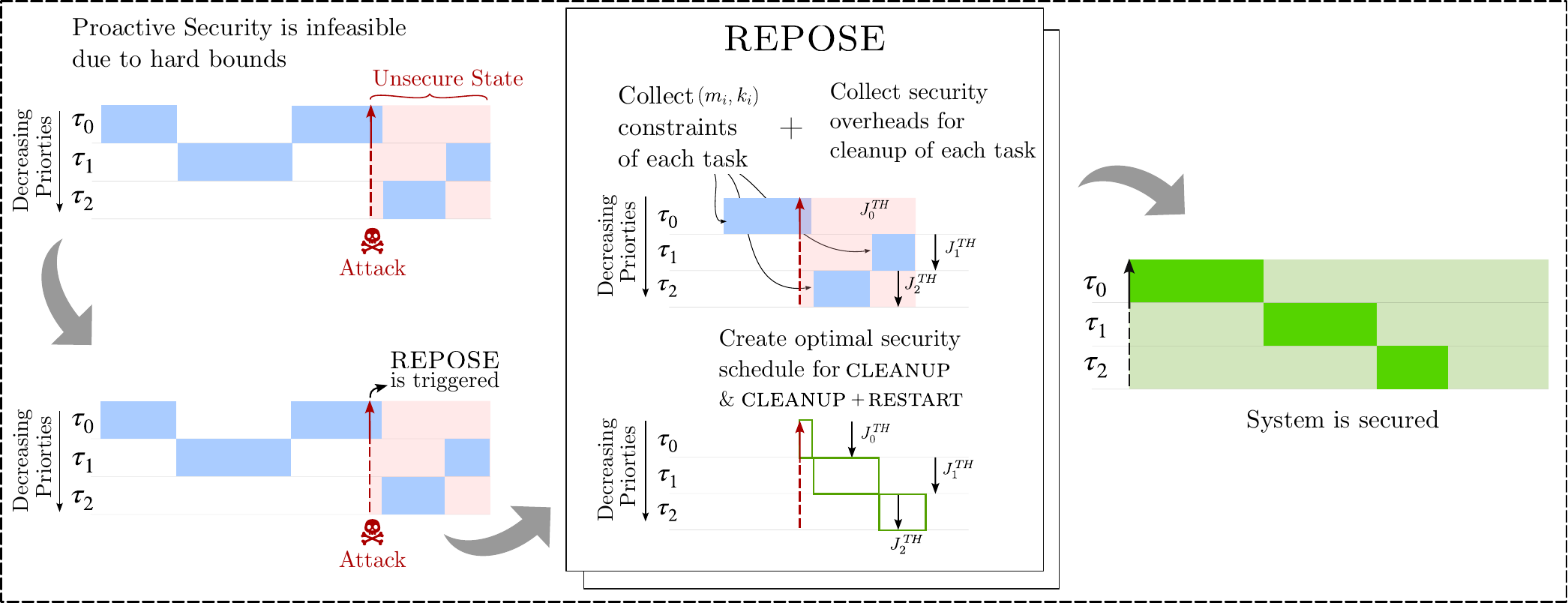}
    \caption{High-level overview of security integration in real-time control tasks using \repose.  Tight system bounds and high utilization prevent the deployment of proactive security mechanisms, rendering the system insecure for an extended period (left block in the figure). \repose\ uses an online security trigger to execute a two-phased security execution. First, \repose\ effectively calculates the price of security that the system can tolerate, then \repose\ creates an optimal security schedule that combines \cleanup\, \onlyrestart, and \restart\ mechanisms, minimizing the cost of security (middle block). These security operations are then executed in the system according to the \repose\ schedule (right block).}
    \label{fig:teaser}
\end{figure*}

Prior research on integrating security in real-time edge devices has revolved around two main approaches: design-time security parameter selection and runtime detection. Design-time techniques such as Contego~\cite{hasan2022beyond} and SchedGuard~\cite{chen2021schedguard} either prolong task execution time by incorporating security mechanisms or rely on opportunistic execution, which may be infeasible in high-utilization systems. Imposing a constant security overhead on each iteration of the task, as considered in some work~\cite{baruah2022security,baruah2023scheduling}, can significantly affect the feasibility of the system. In the area of runtime anomaly detection, prior work investigated anomaly detection and preventive actions (e.g., task termination) based on execution behavior~\cite{hamad2018prediction}. However, these online mechanisms often neglect the critical aspects of resource cleanup and task recovery after an attack. Moreover, existing solutions are not designed for weakly-hard RT-CPS. \repose\ aims to augment the limitations of existing work. The proposed \repose\ framework comprises two primary components: \ca a reactive security trigger that responds to the online behavior of tasks and \cb a weakly-hard control system model. Both techniques have been examined \textit{individually} in prior research~\cite{vreman2021stability,salamun_weakly_2023,pazzaglia_dmac_2019}. However, integrating these components allows \repose\ to establish \textit{analytical bounds} and develop an online scheduling algorithm that incorporates a security mechanism into weakly-hard RT-CPS.

An overview of the \repose\ mechanism is illustrated in Figure~\ref{fig:teaser}.  The security mechanism employed by \repose\ operates in two distinct phases. When a security anomaly is detected, \repose\ is activated by the scheduler. The \textbf{first phase} involves determining the additional overhead that the system can accommodate. In the \textbf{second phase}, \repose\ creates an optimal schedule for the security operations based on the tolerable overhead of the system. The security operations performed in the second phase of \repose\ are categorized into three operations: \cleanup\, \onlyrestart, and \restart\ as shown in Figure~\ref{fig:teaser}.
In \cleanup\ mode, the system resources allocated to the task are flushed or invalidated by \repose, thereby allowing a fresh set of resources to be utilized in the subsequent iteration of the task. \onlyrestart\ mode discards the results from the current execution of the task and restarts the task to recalculate everything. In \restart\ mode, \repose\ first carries out the cleanup operation and then restarts the task, provided that this restart can be completed without violating weakly-hard requirements (i.e., $(m,k)$-constraints of each task). The use of $(m,k)$ constraints in this paper makes \repose\ applicable to a wider range of systems, including ``hard'' RT-CPS, which is a special case of weakly-hard system with $(m,k)$ constrained defined as $(0,1)$, meaning that the system can tolerate $0$ deadline misses on each invocation of the task.

In this paper, we make the following contributions:
\begin{itemize}
    \item A new two-phase security mechanism---\repose---that bounds the overheads and characterizes the ``price'' of integrating security into weakly-hard RT-CPS (Section~\ref{sec:repose}).
     \item An analytical framework for integrating online security in edge devices by extending the control delay to account for security overhead added at an arbitrary time (Section~\ref{sec:hard_constraint}).
    \item An online task selection algorithm to schedule security cleanups based on the feasibility of such operations (Section~\ref{sec:alg_design}).

\end{itemize}

We characterize security actions in RT-CPS using a classical control model and evaluate the price of security across various system parameters under weakly-hard constraints (Section~\ref{sec:eval}). In particular, we show that the analytical framework of \repose\ provides an \textit{exact measure of the security overhead} added to the system, and how that impacts the overall system performance. We also observe that weakly-hard constraints are a better-suited model for handling security violations, while achieving better performance than hard real-time constraints for security monitoring tasks, as assumed in prior work. We also show that the priority-based algorithm used by \repose\ reduces the impact of the security scheduling algorithm on lower-priority tasks, making it suitable for use with classical scheduling algorithms like Rate-Monotonic Scheduling~\cite{lehoczky1989rate} that are known to be vulnerable to side-channel attacks~\cite{nasri2019pitfalls}.

\section{Threat Model}\label{sec:threat_model}

We consider an adversary capable of taking control of a subset of control tasks and rendering the system unsafe---an assumption typical in RT-CPS security research~\cite{hasan2024sok}. For instance,  an attacker can gain information about the system through side-channel analysis and exploit it to execute active attacks, such as DoS attacks. These side-channel attacks target system resources, such as caches~\cite {trilla2018cache}, or observe the temporal properties of real-time tasks~\cite{yoon2021blinder} to compromise the integrity of the system. Using this side-channel information, an adversary can further compromise availability. 

In the context of real-time scheduling, side-channel attacks exploit the temporal properties of a \emph{victim} task. These attacks are broadly classified into the following three categories~\cite{nasri2019pitfalls}.
\begin{enumerate}
    \item \textbf{Anterior attack:} The attacker manipulates the memory, cache, or other system resources \emph{before} the victim task accesses it.
    \item  \textbf{Posterior attack:} The attacker manipulates the system resource \emph{after} the victim task accesses it.
        \item \textbf{Pincer attack:} The adversary manipulates the system resource \emph{before and after} the task accesses it.
    
\end{enumerate}

Among these three broad categories, \repose\ addresses Posterior and Pincer attacks by cleaning up system resources \emph{after} a victim task has utilized them. Additionally, \repose\ initiates a new instance of the victim task following this cleanup, allowing the prior execution to be discarded. It is important to note that Anterior attacks are not relevant to \repose\ as the resource cleanup occurs after the task's execution. In contrast, Anterior attacks aim to target resources before a task accesses them. Because \repose\ is based on observing a task's anomalous behavior, cleaning up resources is not possible before the task first accesses them. Therefore, \repose\ specifically targets Posterior and Pincer attacks.

\repose\ primarily focuses on side-channel attacks that exploit temporal and spatial properties. However, \repose\ is applicable in any resource-based attacks that can be detected through an IDS. We assume that the scheduler can observe the impact of an attack on a task's execution. Such properties may include temporal factors, such as execution time, or spatial factors, such as an unusual increase in memory usage. Our assumptions of tasks' behavior observation are consistent with the real-time literature on online anomaly detection~\cite{hamad2025enhancing,hamad2018prediction}.

\section{System  Model}\label{sec:sys_model}

\subsection{Real-Time Task Model}

We consider a unicore RT-CPS with $\mathcal{N}$ sporadic tasks. Each task $\tau_i$ is characterized by the tuple $(C_i, T_i, J_i, J^{TH}_i, \lambda_i)$, where $C_i$ is the worst-case execution time (WCET), $T_i$ is the task period or the minimum interarrival time of two consecutive instances of $\tau_i$. Each task has a weakly-hard requirement, $\lambda_i = (m_i, k_i)$, meaning that the system is considered safe if, in $k_i$ consecutive invocations, at least $m_i$ instances meet the deadline. The parameters $J_i$ and $J^{TH}$ are the measured control performance and the performance threshold, respectively, as described in detail in Section~\ref{sec:ctrl_sys}.

We assume that a scheduler knows the duration of a task that has executed on the processor and the amount of time remaining, based on WCET estimations, using the task's control block (TCB). This information is commonly available in existing sporadic task scheduler implementations, such as the POSIX sporadic server~\cite{stanovich2010defects}. We further assume that a security module is integrated into the scheduler. The scheduler can use the task execution times to determine if a security cleanup is required. Integrating security decisions into the scheduler has been explored previously and has been demonstrated to be a feasible approach for enhancing the security posture of real-time systems~\cite{hasan2016exploring, hasan_contego_2017, hasan2022beyond, baruah2022security,  baruah2023scheduling}. In contrast to prior work, this paper relaxes the hard real-time requirements on security mechanisms, allowing a few tasks to miss their deadlines---albeit with \textit{bounded} misses---in the interest of security. Such system requirements have been extensively studied and analyzed in prior work~\cite{bernat2001weakly, xu2015improved}. We use weakly-hard models to bound the deadline misses the system can tolerate without compromising its physical safety.

Weakly-hard requirements, as defined by $(m,k)$ parameters, can be classified based on the deadline \emph{meet} and \emph{miss} patterns~\cite{bernat2001weakly}. In this work, we consider the weakly-hard constraint $\lambda_i$ to be of type \emph{any $m$ misses}, which means that $\lambda_i$ will be regarded as satisfied if $\tau_i$ misses a maximum of $m_i$ deadlines within $k_i$ consecutive invocations, irrespective of the order of the meets. This type of weakly-hard system has been used in prior work to determine safe operation of control systems~\cite{hobbs2024quantitative}.  Note that a hard-constrained system is a special case of a weakly-hard system, where $m = 0$ at all times, \ie, the system cannot tolerate any deadline misses. 

For each $\tau_i$, the deadline meet and miss pattern is stored in the form of a binary string denoted by $\mu_i$. Each deadline miss is stored as a \verb|1| and each deadline miss is stored as a \verb|0| in $\mu_i$. For example, if $\tau_i$ meets three deadlines, followed by a deadline miss for $k_i = 4$ consecutive invocations, then $\mu_i$ for those $k_i$ invocations will be \verb|"1110"|. The scheduler uses the $\mu_i$ associated with each task to determine whether a future deadline miss is feasible. We assume that the miss pattern for a target system is known at design time, for instance, using techniques developed in earlier work~\cite{bernat2001weakly,hobbs2024quantitative,moyano2025efficient}.

\subsection{Plant Model}\label{sec:ctrl_sys}

We now present the plant model and recap the schedulability conditions derived in prior work~\cite{bini2008delay,hasan2022beyond}. We consider a control system with a linear-quadratic Gaussian framework~\cite{cervin2002feedback,xu2014response}, where each plant, $P_i$ can be defined using the following equations:
\begin{align}
    \dot x_i(t) &= A_ix_i(t) + B_iu_i(t) + v_i(t) \label{eq:state_rate}\\
    y_i(t_k) &= C_ix_i(t_k) + e_i(t_k), \label{eq:control_output}
\end{align}
where $x_i$ is the plant state, $u_i$ is the control input, and $v_i$ is plant disturbance. $y_i(t)$ is the output measured at time $t$ with a measurement noise  $e_i(t)$. The control performance is defined as follows~\cite{bini2008delay,hasan2022beyond}:
\begin{align}
    J_i = \lim_{t \to \infty} \frac{1}{t} \mathbb{E} \Bigg\{ \int_0^t ( y^2_i(\sigma_i) + \rho_iu^2_i(t))d\sigma \Bigg\}\label{eq:control_perf},
\end{align}
In the above equation,
    $\mathbb{E}$ is the expectation operation, 
    $\rho_i$ is a designer-provided weighting factor and
    $J_i$ is a measure of performance loss when the response time is different from the design-time assumption.

For a given period $T_i$ and control delay $\Delta_i$ (e.g., due to execution of the task and preemption by other high-priority tasks), the control cost in Equation~\ref{eq:control_perf} can be approximated by the following linear
function~\cite{bini2008delay,hasan2022beyond}:
\begin{align}
    J_i = \alpha_iT_i + \beta_i\Delta_i, \label{eq:cost_func}
\end{align}
where $\alpha_i$ and $\beta_i$ are the weighting factors used for linearization and can be obtained
from the physical properties of $P_i$. Note that Equation~\ref{eq:cost_func} is generalizable as it depends on $\Delta_i$ and $T_i$ for each task. Hence, the exact cost model shown in Equation~\ref{eq:control_perf} can be modified to fit different plant models based on the system designer's needs. Generalizing the control performance in Equation~\ref{eq:cost_func} enables us to study the impact of security techniques on the control delay, which is critical to the real-time performance of the plant.

We consider the control task to be feasible and useful if the following condition is met:
\begin{align}
    \alpha_iT_i + \beta_i\Delta_i \leq J^{TH}_i.\label{eq:threshold}
\end{align}
The control delay $\Delta_i$ can be calculated using classic response-time analysis using the concept of \emph{busy period} or \emph{busy window}~\cite{lehoczky1990fixed}:
\begin{align}
\delta_i(q) = (q + 1) C_i +
\sum_{\tau_j \in hp(\tau_i)}
\left\lceil \frac{\delta_i(q)}{T_j} \right\rceil C_j,\label{eq:resp_time}
\end{align}
where $q$ is the index of each job in the busy window and $\tau_j \in hp(\tau_i)$ returns all the tasks in the system that have a higher priority than that of $\tau_i$. The recurrence in Equation~\ref{eq:resp_time} can be solved by taking the base conditions as $\delta_i(q)^0 = (q + 1) C_i$ and $\delta_i(0)^0 = C_i.$ The recurrence will terminate when $\delta_i(q)^k = \delta_i(q)^{k-1}.$ Hence, the control delay can be expressed as:
\begin{align}
    \Delta_i(q) = \delta_i(q) - q T_i.
\end{align}
Then, the worst-case delay will be:
\begin{align}
    \Delta_i = \max_{\forall q = \{0, 1, \ldots\}} \{ \Delta_i(q) \}.~\label{eq:ctrl_delay}
\end{align}

\subsubsection{Interplay Between Control Updates and Security Requirements}

Based on our characterization of the control system in Equations~\ref{eq:state_rate} and~\ref{eq:control_output}, the output and the state of the task mainly depend on the state of the task and control input provided at time $t$, \ie $x_i(t)$ and $u_i(t)$ respectively. From the control task perspective, $u_i(t)$ is generally provided to the task through some buffered input, for example, a temperature sensor buffer. Such control is often stored in system resources such as shared cache or shared buffers. Prior research~\cite{chen2021schedguard,chen2023schedguard++,liu2015last} has shown that such shared resources are vulnerable to resource-based attacks, such as a side-channel or a data injection. Such side-channel attacks are proven to be practical and feasible in both general-purpose systems~\cite{liu2015last} as well as RT-CPS~\cite{mohan2014real}. To recover from such an attack,~\repose\ uses a~\cleanup\ mechanism to clear a task's system resources and mitigate the impact of side-channel attack on the shared resource.

The state of the task $x_i(t)$ depends on the local variables, any calculations, and the state generated from the control input at time $t$. As shown in Equation~\ref{eq:ctrl_delay}, the task calculates this state during $\Delta_i$. In case of an attack, the state of the task generally changes, causing the system to give unreliable output~\cite{chen_novel_2019}. Consistent with the existing literature~\cite{abdi2016resecure,banerjee2022secure}, we assume each task is stateless, meaning that the system has no memory of the prior states and recalculates the current state based on the given input. \repose\ re-instantiates the state for the system by discarding the current state and using a task-level~\onlyrestart\ mechanism. Through online real-time calculation of the tolerable delays (see Section~\ref{sec:hard_constraint}), \repose\ calculates the right combination of~\cleanup,~\onlyrestart, or ~\restart\ mechanisms for each task to minimize the \emph{price of security} to the system.

\subsection{Security Overhead}\label{sec:cleanup_overhead}

The security operation is responsible for cleaning, restarting, or both operations based on the available system budget and the system resources used by a task. For each task $\tau_i$, we denote the security overhead as $\kappa_i$. For \cleanup\ operation, the overhead depends on the total amount of system resources allocated to each task. The exact value of the overhead can be calculated by the system designer. In the rest of the text, we assume that the cleanup overhead is a fraction $\psi_i$ of the WCET of the task. Hence, $\kappa_i$ for the cleanup operation is $\psi_iC_i$. In case of \onlyrestart---unlike \cleanup\ operation---the overhead of task initialization is already a part of the $C_i$. Hence, the total overhead in~\onlyrestart\ will be the task's execution time, $C_i$. For a combined~\restart\ operation, the total overhead of the security operation is $\kappa_i = C(1+\psi_i)$.

\repose\ first checks the feasibility of~\cleanup,~\onlyrestart, and~\restart\ operations, and creates the optimal security operation schedule to minimize the price of security on the system. To minimize the impact of the security on high-priority tasks, \repose\ executes \cleanup\ part of the security operation at the highest priority, but after the task is re-initialized, the task executes in its normal priority and can be preempted by other higher-priority tasks.

\subsection{Online Security Detection}
The security operation is triggered based on the execution behavior of the control tasks. If any anomaly is detected, the scheduler triggers \repose\ and schedules security operations accordingly. We consider an anomaly detection system that uses task execution behavior to flag tasks as \emph{victim} tasks, i.e., those potentially under attack by an adversary. The execution behavior of tasks can exhibit \emph{temporal} patterns, such as prolonged execution and frequent exceedance of the average execution time. Attack artifacts can also be \emph{spatial}, such as a sudden increase in memory usage. 

We assume that the detection mechanism functions correctly and triggers \repose\ whenever a security event occurs. The exact detection operation is beyond the scope of this study. Many techniques have been introduced in the literature for detecting anomalous behavior in RT-CPS~\cite{mitchell2014survey,asadi2013run}, which can be integrated with \repose\ without loss of generality. For instance, the \emph{RedZone} principle~\cite{hamad2018prediction} is one technique that leverages real-time properties to detect whether a task may be compromised. In RedZone, a task is flagged as anomalous if it executes beyond the pre-calculated execution bounds, even if no deadline miss is observed. In this paper, we use the terms ``RedZone'' and ``security trigger'' to refer to a security event detected by an online detection mechanism. Section~\ref{sec:repose} presents the online feasibility check and security scheduling operations in detail.

\subsection{Security Cleanup Mechanism}\label{sec:cleanup_mech}

We now present a \cleanup\ mechanism that can be used with \repose. However, note that the following cache-based \cleanup\ mechanism is \emph{one potential mechanism} to demonstrate \repose\ operation. \repose\ can be used with \emph{any such mechanism} as long as it has bounded overhead per task.

\begin{algorithm}[!t]
\caption{\cleanup\ Operation}
\label{alg:cleanup_mechanism}
\small
\begin{algorithmic}[1]
\State \textbf{Input:} $
\Gamma_{vic}$\Comment{Set of tasks flagged as under attack} \label{alg_cleanup:inupt}
\State \textbf{Output:} Cleanup the cache related to the tasks
\State $\chi$ \Comment{Resource (Cache) block size}\label{alg_cleanup:chi}
\State $\Phi(\cdot)$ \Comment{Cleanup (Cache flush) Operation}\label{alg_cleanup:phi}
\For{$\tau_i \in \Gamma_{vic}$}
\State $\mathcal{S}_i(\epsilon, \omega),  \gets$ Stack with start address $\epsilon_i$, and size $\omega_i$ \label{alg_cleanup:stack_decl}
\State  $\epsilon'_i \gets \Big\lfloor\frac{\epsilon_i} {\chi}\Big\rfloor \chi$ \Comment{Start address aligned to resource block}\label{alg_cleanup:start_align}
\State  $\omega'_i \gets \Big\lceil\frac{\omega_i}{\chi}\Big\rceil \chi$\Comment{Size aligned to resource block}\label{alg_cleanup:size_align}
\State $\Phi\big(\mathcal{S}_i(\epsilon'_i, \omega'_i)\big)$\Comment{Flush aligned resource}\label{alg_cleanup:cleanup}
\EndFor
\end{algorithmic}
\end{algorithm}

\begin{figure*}
    \begin{subfigure}[t]{0.5\textwidth}
     \includegraphics[width=0.8\textwidth]{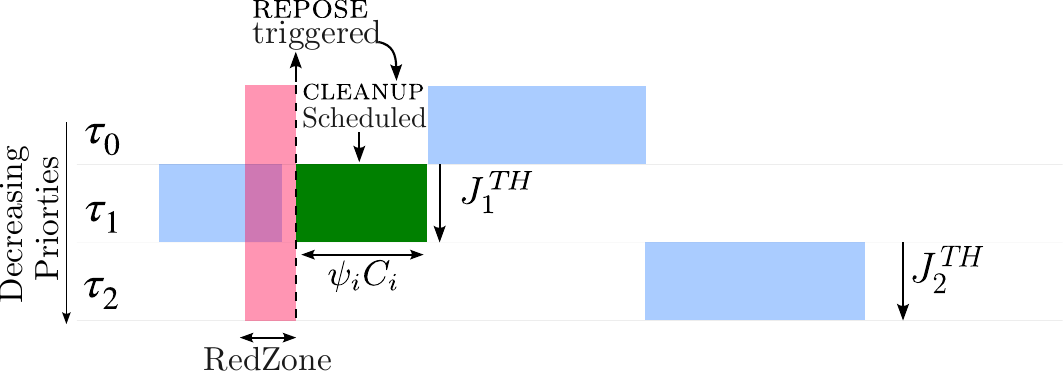}
     \caption{Feasible \cleanup.}

    \label{fig:sec_cleanup}
    \end{subfigure}
    \hfill
   \begin{subfigure}[t]{0.5\textwidth}
         \includegraphics[width=0.8\textwidth]{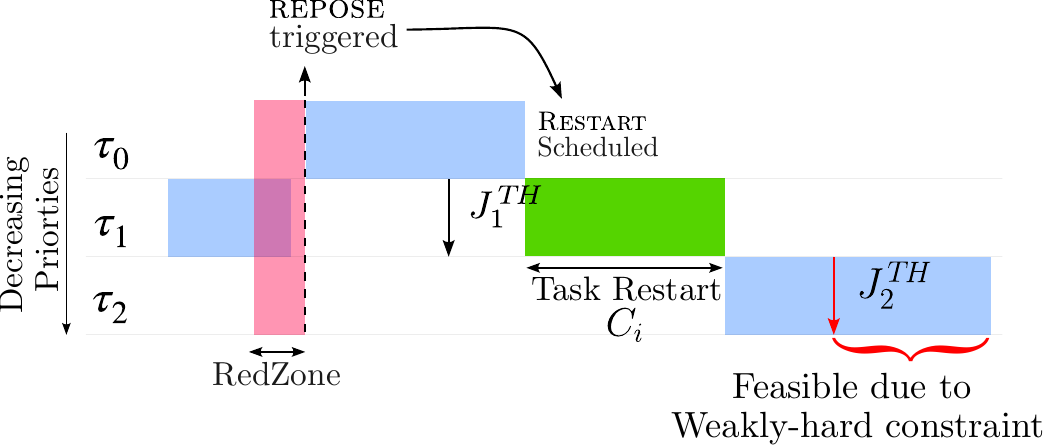}
    \caption{Feasible \onlyrestart.}

    \label{fig:sec_onlyrestart}
   \end{subfigure}
   \par
   \begin{subfigure}[t]{0.5\textwidth}
    \includegraphics[width=0.8\textwidth]{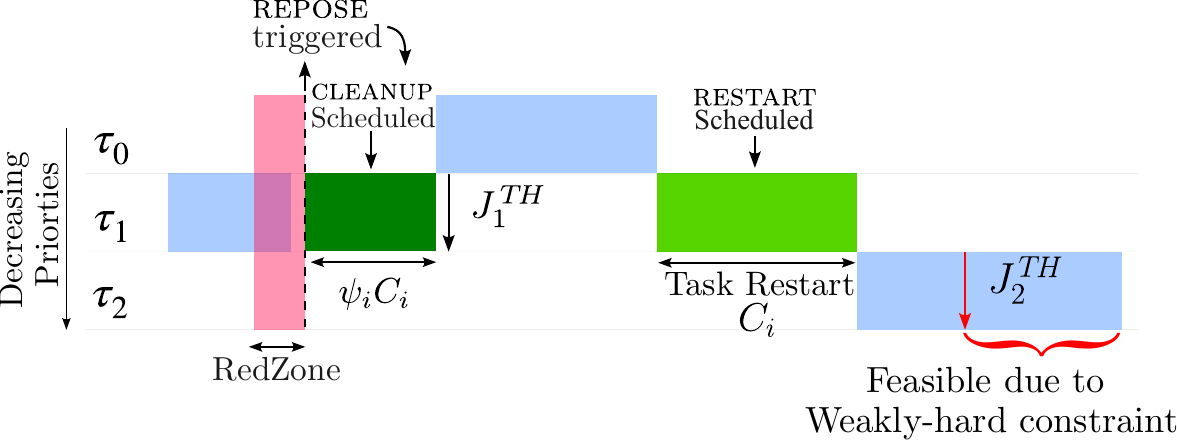}
    \caption{Feasible \restart.}

    \label{fig:sec_restart}
    \end{subfigure}
    \hfill
    \begin{subfigure}[t]{0.5\textwidth}
    \includegraphics[width=0.8\textwidth]{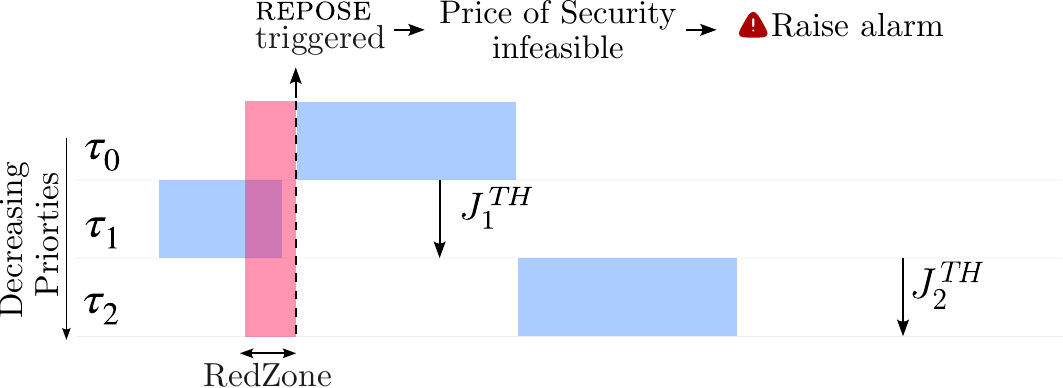}
    \caption{Any security operation is infeasible.}

    \label{fig:sec_inf}
    \end{subfigure}
    \caption{Different Cases of Security Operations. \ca \cleanup\ is feasible even under hard constraints. Security operations (\cleanup) are triggered when the task enters a RedZone. \cb \onlyrestart\ is feasible under weakly-hard constraints. \cc \restart\ is feasible under weakly-hard constraints. This case provides the highest security at the highest cost to the system. \cd No added security operation is feasible---this case can emerge if the security incidents are too frequent. To avoid a DoS possibility, \repose\ immediately raises an alarm without preempting the executing tasks of the system. However, after the point of alarm, the system states are unreliable due to an attack.}
\end{figure*}

Algorithm~\ref{alg:cleanup_mechanism} shows a generic mechanism for a cleanup operation after a task is identified as vulnerable. Such an algorithm can be implemented on most commercial-off-the-shelf (COTS) RTOS, such as RTEMS~\cite{bloom2020real}, which is commonly used in safety-critical control systems, including the EPICS middleware used for experimental physics~\cite{lee2016distributed}. We next describe how Algorithm~\ref{alg:cleanup_mechanism} can be implemented in RTEMS, treating the cache as the target system resource that requires cleanup.

Line~\ref{alg_cleanup:inupt} takes the list of victim tasks, $\Gamma^t_{vic}$, as input. These tasks have been marked as victim tasks using the RedZone principle. Each task has a thread control block (TCB) that stores the task context required by the scheduler to schedule tasks correctly.  The vulnerable tasks are identified using the task ID stored in the TCB.
Lines~\ref{alg_cleanup:chi}-\ref{alg_cleanup:phi} are variables used to indicate the resource block size and an OS cleanup operation. The resource here is the cache, and the cache line size is a property of the processor architecture. The parameter $\chi$ can be obtained in RTEMS using a standard API call \verb|rtems_cache_get_data_line_size()|.

The cleanup operation, $\Phi(\cdot)$, is typically an OS API call associated with the target resource. In this case, $\Phi(\cdot)$ is a cache flush or invalidate operation. In RTEMS, such a flush operation can be performed using \verb|rtems_cache_flush_multiple_data_lines| or \verb|rtems_cache_invalidate_multiple_data_lines|. These cache flush operations are common in most embedded OS, and are commonly used for cache management protocols like cache coherence~\cite{gracioli2015survey}. Similar cache cleaning mechanisms have also been used in real-time security contexts~\cite{mohan2014real,pellizzoni2015generalized}. Lines~\ref{alg_cleanup:start_align} and~\ref{alg_cleanup:size_align} align the start address and the size of the total cleanup resources to be flushed. Note that in Line~\ref{alg_cleanup:size_align} we round up the resource cleanup to ensure the entire region is flushed. The added overhead is due to the security operation, which impacts all the tasks in the system. The analysis presented next accounts for this overhead.

\section{REPOSE Design and Analysis}\label{sec:repose}

\repose\ is an online security mechanism that utilizes known task parameters to trigger a security operation. \repose\ uses \textit{control delay bounds} to check the feasibility of any security operation before scheduling them. The security operations \cleanup, \onlyrestart, and \restart\ target the system resources allocated to a task and use available resource flush and invalidation operations to clean up the system resources used by the victim task. Since \repose\ does the cleaning up and invalidation of the resource used by a task, it mitigates the possibility of an adversary accessing the same resources after the task's execution to breach the system integrity through a posterior or pincer attack identified in prior work~\cite{nasri2019pitfalls}.

Figure~\ref{fig:sec_cleanup} illustrates a time window where \cleanup\ is triggered by a RedZone-based online detection. Here, the overhead $\psi_iC_i$ is feasible, but performing \restart\ will exceed the performance cost threshold, $J^{TH}$, making the restart operation infeasible under the hard constraint. Similarly, in Figure~\ref{fig:sec_onlyrestart}, only \onlyrestart\ is feasible without a cleanup operation.  Figure~\ref{fig:sec_restart} shows a scenario where a complete~\restart\ is feasible even if the total overhead goes beyond $J_i^{TH}$ due to the weakly-hard bounds of the system. In Figure~\ref{fig:sec_inf}, we see a case where neither of the security operations will be feasible under hard constraints. For the case presented in Figure~\ref{fig:sec_inf}, \repose\ will \emph{notify} the admin of a potential security incident and allow the admin to decide how to handle it on a case-by-case basis. The following section formally derives the feasibility conditions.

\subsection{Feasibility Analysis}\label{sec:hard_constraint}

We now present the task selection mechanism of \repose\ under the weakly-hard constraints.  When appropriate, we compare/contrast \repose\ with hard constraints where the system cannot tolerate any deadline miss. For instance, Figure~\ref{fig:sec_cleanup} illustrates a scenario when a~\cleanup\ operation is feasible under hard real-time constraints (and hence, for weakly-hard systems as well).

The security operations of~\repose\ are triggered by an online anomaly checker (such as RedZone~\cite{hamad2018prediction}). If an anomaly is detected at time $t$, the scheduler stores the context of the flagged tasks in a victim task set $\Gamma_{v}$ and calculates if a security operation is possible without jeopardizing system safety.

Let $\zeta_i(t)$ be the release time of an instance of $\tau_i$, such that $\zeta_i(t) <t < \zeta_i(t) + J^{TH}_i$. Then $\zeta_i(t)$ can be defined as:
\begin{align}
    \zeta_i(t) = \Bigg\lfloor\frac{t}{T_i}\Bigg\rfloor T_i.
\end{align}
Hence, the execution window of $\tau_i$ containing $t$ is $[\zeta_i(t), \zeta_i(t) + J^{TH})$. Lemma~\ref{lemma:perf_spec_window} shows the control performance for a given execution window that contains the time $t$. 

\begin{lemma}\label{lemma:perf_spec_window}
For a given time $t \in [\zeta_i(t), \zeta_i(t) + J^{TH}_i)$, $\tau_i$ meets the deadline if the control cost meets the following constraint:

\begin{align}
   J_i(t) = \alpha_i T_i + \beta_i\Delta_i(t) \leq J^{TH}_i,
\text{where }
    \Delta_i(t) = \Delta_i - \zeta_i(t).\label{eq:delta_window_t}
\end{align}
\begin{IEEEproof}

This result is an extension of Equation~\ref{eq:threshold} where we consider control delay faced by $\tau_i$ during the time window that contains $t$. Note that the control delay $\Delta_i$ is the worst-case control delay from Equation~\ref{eq:ctrl_delay}. Considering the worst-case bound to calculate the control delay at $t$ ensures that Equation~\ref{eq:delta_window_t} is generalizable for all instances of $\tau_i$. Hence, Equation~\ref{eq:delta_window_t} provides the control cost of any task $\tau_i$ whose execution window contains $t$. The relative control delay is derived using Equation~\ref{eq:delta_window_t}, considering the release of the instance of $\tau_i$ released in the execution window immediately preceding $t$.
\end{IEEEproof}
\end{lemma}

The performance cost calculated in Equation~\ref{eq:delta_window_t} gives the cost of the full window of execution that contains time $t$. At an arbitrary time $t$, the feasibility of any added security overhead would depend on how much execution time is \emph{remaining} for $\tau_i$ at time $t$. Lemma~\ref{lemma:theta_define} illustrates the modification in the control delay equation required to calculate the remaining control $\hat\Delta_i(t)$ at time $t$.

\begin{figure}
    \centering
    \includegraphics[width=\columnwidth]{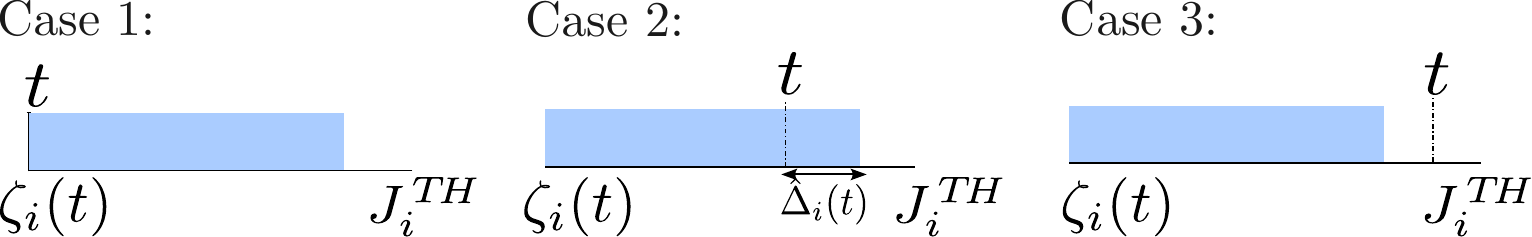}
    \caption{An arbitrary point $t$ can be placed in three cases relative to the execution window of $\tau_i$. In Cases 1 and 2, $t$ does not interfere with the execution of $\tau_i$. In Case 2, at point $t$, there is still some execution time left for $\tau_i$ to finish its current execution. Compared to the prior cases, Case $3$ shows a situation where the complete execution of $\tau_i$ finishes before $t$.}
    \label{fig:Conditions_t}
\end{figure}

To calculate the remaining control delay of $\tau_i$, we should consider three cases:
\begin{enumerate}
    \item $t=\zeta_i(t)$\label{case:equal} $\Rightarrow$ when $t$ is aligned with the release of $\tau_i$,
    \item $\zeta_i(t) < t < \zeta_i + \Delta_i$\label{case:mid} $\Rightarrow$ when $t$ is in the middle of an incomplete execution of $\tau_i$, and
    \item $t> \zeta_i(t) + \Delta_i(t)$ \label{case:end}  $\Rightarrow$ when $t$ comes after $\tau_i$ finishes execution.
\end{enumerate}

These three conditionals are illustrated in Figure~\ref{fig:Conditions_t}. In Case~\ref{case:equal}, the control delay equation can be obtained using Equation~\ref{eq:delta_window_t}. For Case~\ref{case:end}, since the task has already finished execution, the remaining control delay would be $0$. In Cases~\ref{case:equal} and~\ref{case:end}, the control task has either not started yet, or has already completed its execution. In Case~\ref{case:mid}, the control task has only partially executed and the amount of remaining control delay depends on the amount of time $\tau_i$ has already spent on the processor, the amount of time remaining to finish the rest of the execution without exceeding $J_i^{TH}$, and the interference from other tasks in the system during the time window $[t, \zeta_i(t) + J_i^{TH})$. Lemma~\ref{lemma:theta_define} formally defines these three different cases and derives the equation for $\hat\Delta_i(t)$.

\begin{lemma}\label{lemma:theta_define}
For a given time $t \in [\zeta_i(t), \zeta_i(t) + J^{TH}_i)$, the remaining control cost of a partially executed instance of $\tau_i$ is calculated as follows:
\begin{align}
      \hat J_i(t) = \alpha_i T_i + \beta_i\hat\Delta_i(t) \leq J^{TH}_i - (t - \zeta_i(t)),\label{eq:partial_perf}
\end{align}
where
\begin{align}
    \hat\Delta_i(t) = 
    \begin{cases}
        \Delta_i - \zeta_i(t), & t=\zeta_i(t) \\
        \theta_i(t), & \zeta_i(t) < t < \zeta_i + \Delta_i \\
        0, & t> \zeta_i(t) + \Delta_i(t)
    \end{cases}
\end{align}
and 
\begin{align}\label{eq:theta_define}
    \theta_i(t) = &\hat C_i(t) + \sum_{\substack{\tau_j \in hp(\tau_i)\\\tau_j\in \Gamma^t}}\hat C_j(t)  
    + \\
    &\sum_{\tau_j \in hp(\tau_i)}{\Bigg(\Bigg\lceil \frac{\zeta_i(t) + J^{TH}_i}{T_j}\Bigg \rceil - \Bigg\lfloor \frac{t}{T_j} \Bigg\rfloor}\Bigg)C_j.
\end{align} 
Here, $\hat C_i(t)$ is the remaining execution of $\tau_i$ at time $t$. $\Gamma^t$ is the set of all tasks in the system that are in blocked or execution state at time $t$. 
\begin{IEEEproof}
    The term $\sum_{\tau_j \in hp(\tau_i)}{\Big(\Big\lceil \frac{\zeta_i(t) + J^{TH}_i}{T_j}\Big \rceil - \Big\lfloor \frac{t}{T_j} \Big\rfloor}\Big)C_j$ provides the interference from the execution of all the higher priority tasks that were released during the time window $[\zeta_i(t), \zeta_i(t)+J^{TH}_i)$, and subtracting the execution overheads from all the executions before $t$. The ceil operation in the first term and the floor operation in the second term accommodate the back-to-back hit problem~\cite{audsley1993applying}.
\end{IEEEproof}
\end{lemma}

As defined in Section~\ref{sec:cleanup_overhead}, the overhead for security operations can be characterized as:
\begin{align}
    \kappa_i = 
    \begin{cases}
        \psi_iC_i, & \text{\textit{Cleanup}}\\
        C_i, & \text{\textit{Restart}}\\
        C_i(1 + \psi_i),& \text{\textit{Cleanup+Restart}}.
    \end{cases}
\end{align}

\begin{theorem}\label{theorem:delta_security}
If there are a set of victim tasks $\Gamma_{vic}$, $\Gamma^{c}_{vic} \subset \Gamma_{vic}$ tasks are scheduled for \cleanup\ operation, $\Gamma^{r}_{vic} \subset \Gamma_{vic}$ tasks are scheduled for \onlyrestart, and $\Gamma^{cr}_{vic} \subset \Gamma_{vic}$ scheduled for \restart\ then the total remaining control delay can be calculated as follows:
\begin{equation}
\label{eq:delta_sec} 
\begin{aligned}
\hat\Delta_i(t, \Gamma^c_{vic}, \Gamma^r_{vic}, \Gamma^{cr}_{vic}) = \hat C_i(t) + 
\sum_{\substack{\tau_j \in hp(\tau_i) \\\tau_j\in \Gamma^t}}{\hat C_j} + \sum_{\substack{\tau_j \in hp(i)\\\tau_j\in \Gamma^{r}_{vic}}}C_j\\
    + \sum_{\tau_j \in hp(\tau_i)}{\Bigg(\Bigg\lceil \frac{\zeta_i + J^{TH}_i}{T_j}\Bigg \rceil - \Bigg\lfloor \frac{t}{T_j} \Bigg\rfloor}\Bigg)C_j 
    + \sum_{\substack{\tau_j \in hp(i)\\\tau_j\in \Gamma^{cr}_{vic}}}C_j(1 + \psi_j) \\
    +\sum_{\substack{\tau_j\in \Gamma^{c}_{vic}}}\psi_jC_j.
\end{aligned}
\end{equation}

For $t= \zeta_i(t)$, $\hat C_i(t) = C_i$ and Equation~\ref{eq:delta_sec} is equivalent to Equation~\ref{eq:theta_define}, if $\Gamma^c_{vic} = \emptyset\And \Gamma^r_{vic} = \emptyset \And \Gamma^{cr}_{vic}=\emptyset$. 

The security operation meets the deadline under the following constraint:
\begin{align}
    \hat\Delta_i(t, \Gamma^c_{vic}, \Gamma^r_{vic}, \Gamma^{cr}_{vic})^{TH} \leq \frac{J^{TH}_i - \alpha_iT_i - (t - \zeta_i(t))}{\beta_i}. \label{eq:delta_mod}
\end{align}

\begin{IEEEproof}
 Equation~\ref{eq:delta_sec} is derived from extending Lemma~\ref{lemma:theta_define} to include the overhead terms for the security operations. The \cleanup\ affects all the tasks in the system to ensure that no malicious tasks can execute before the \cleanup\ is complete. Hence, the \cleanup\ overhead $\psi_jC_j$ affects all tasks in the system. However, in a combined \restart or in case of only \onlyrestart, the added overhead of re-instantiation, $C_j$, for the restarted task will only affect other tasks if $\tau_i$ has a higher priority. Hence, combining the security overhead and the partial control delay at time $t$ in Equation~\ref{eq:delta_sec} provides the total control delay that will be faced by the current execution of a task after the addition of security operations. Equation~\ref{eq:delta_mod} provides the upper bound of the control delay. The result can be derived from rearranging Equation~\ref{eq:partial_perf}. Hence, if the security overhead follows the constraint given in Equation~\ref{eq:delta_mod}, $\tau_i$ will not miss a deadline. 
\end{IEEEproof}
\end{theorem}

We now introduce techniques to verify feasibility before scheduling security and restart operations. 
Lemma~\ref{lemma:feasibility} formally presents the feasibility conditions.

\begin{lemma}\label{lemma:feasibility}
    For a task $\tau_i$ with weakly-hard constraint $(m,k)$ and a $\mu$ string $\mu_i(t)$ at time $t$, let the feasibility of the added security overhead be a binary function $\Omega_i(t, \Gamma^c_{vic}, \Gamma^r_{vic})$, such that $\Omega_i(\cdot)$ is true if the security overhead is feasible. Let $\eta(\mu_i)$ return the hamming weight of the binary string $\mu_i$. Then, the feasibility of the security operation can be determined as follows:
    \begin{align}\label{eq:omega_define}
        \Omega_i(t, \Gamma^c_{vic}, \Gamma^r_{vic}, \Gamma^{cr}_{vic}) =
        \begin{cases}
            1, & \hat\Delta_i(\cdot)^{TH} \leq \frac{J^{TH}_i - \alpha_iT_i - (t - \zeta_i(t))}{\beta_i}\\
            1, & m_i \geq k_i - \eta(\mu_i) + 1\\
            0, &\text{Otherwise}
        \end{cases}
    \end{align}
\begin{IEEEproof}
    The feasibility of an added security depends on two factors: \begin{enumerate}
        \item Control delay constraint in Theorem~\ref{theorem:delta_security} and
        \item The $(m,k)-$ Constraint $\lambda_i$.
    \end{enumerate}
    Hence, $\Omega_i(t, \Gamma^c_{vic}, \Gamma^r_{vic}, \Gamma^{cr}_{vic})$ returns true when either of these above two constraints is met. In second condition, we check $m_i$ against $k_i -\eta(\mu_i) + 1$, the added $1$ is to account for a future deadline miss that can be caused by the addition of the security overhead, since $\Omega(\cdot)$ is calculated by the scheduler at time $t$, \emph{before} the security operations are added into the schedule. Therefore, based on the two constraints mentioned above, we can determine whether adding the required security operations is feasible using Equation~\ref{eq:omega_define}.
\end{IEEEproof}
\end{lemma}

\textbf{Note:} As noted earlier, hard constraints, i.e., the cases when tasks cannot tolerate any deadline misses for any invocation of a task, are a special case of weakly-hard requirements, with $(m,k)$ set to $(0,1)$. For hard systems, \repose\ can be implemented without loss of generality. In this case, the guarantees are enforced by Theorem~\ref{theorem:delta_security} and Lemma~\ref{lemma:feasibility} by setting $m = 0$. Hence, \repose\ is applicable to \textit{both hard and weakly-hard RT-CPS}, although our primary focus is on quantifying the price of security under weakly-hard constraints. Section~\ref{sec:eval_h_vs_wh} presents evaluation results comparing hard and weekly-hard requirements.

\subsection{Algorithm Design}~\label{sec:alg_design}

\begin{algorithm}[t]
\caption{Restarting Tasks Based on Weakly-Hard Constraints}
\label{alg:restart_decision}
\small
\begin{algorithmic}[1]
\State \textbf{Input:} $\Gamma^t, \Gamma^t_{vic} \gets \text{All tasks and all victim tasks at time } t$
\State \textbf{Output: } $\Gamma^{cr}_{vic}, \Gamma^r_{vic}, \Gamma^c_{vic}$, \text{or trigger an ``Alarm'' about other victim tasks}
\State $\Gamma^c_{vic} = \emptyset$ \Comment{Queue of tasks ready for \cleanup}
\State $\Gamma^r_{vic} = \emptyset$ \Comment {Queue of tasks ready \onlyrestart}
\State $\Gamma^{cr}_{vic} = \emptyset$ \Comment {Queue of tasks ready \restart}

\For {$\tau_i \in \Gamma^t_{vic}$}\label{alg2:for_vic} \Comment{Iterating from higher to lower priority}
    \State $\nabla_{c} \gets \emptyset$ \Comment{Set of tasks that can tolerate \cleanup}\label{alg2:nabla_cle}
    \State $\nabla_{r} \gets \emptyset$\Comment{Set of tasks that can tolerate \onlyrestart}\label{alg2:nabla_r}
    \State $\nabla_{cr} \gets \emptyset$\Comment{Set of tasks that can tolerate \restart}\label{alg2:nabla_res}
    \State $\nabla_{inf} \gets \emptyset$ \Comment{Set of tasks for which security is infeasible}\label{alg2:nabla_inf}
    \State $\hat\Gamma^r_{vic} = \Gamma^c_{vic} \cup {\tau_i}$ \Comment{Add $\tau_i$ to temporary \cleanup\ queue}\label{alg2:clea_set}
    \State $\hat\Gamma^r_{vic} = \Gamma^c_{vic} \cup {\tau_i}$ \Comment{Add $\tau_i$ to temporary \onlyrestart\ queue}\label{alg2:onlyres_set}
    \State $\hat\Gamma^{cr}_{vic} = \Gamma^c_{vic} \cup {\tau_i}$ \Comment{Add $\tau_i$ to temporary \restart\ queue}\label{alg2:res_set}
    \For {$\tau_j \in \Gamma^t$}\label{alg2:for_all}
 
    \If {$\Omega_j(t, \Gamma^c_{vic}, \Gamma^r_{vic}, \hat\Gamma^{cr}_{vic}) \And \tau_i \in hp(j)$}\label{alg2:res_feasible}
        \State $\nabla_{cr} = \nabla_{cr} \cup \tau_j$ \Comment{\restart\ is feasible for $\tau_j$}
    \ElsIf { $\Omega_j(t,\Gamma^c_{vic}, \hat\Gamma^r_{vic}, \Gamma^{cr}_{vic})\And \not\Omega_j(t, \Gamma^c_{vic}, \Gamma^r_{vic}, \hat\Gamma^{cr}_{vic}) $}\label{alg2:onlyrestart_feasible}
            \State $\nabla_{r} = \nabla_{r} \cup \tau_j$ \Comment{Only \onlyrestart\ is feasible for $\tau_j$}
    \ElsIf { $\Omega_j(t, \hat\Gamma^c_{vic}, \Gamma^r_{vic}, \Gamma^{cr}_{vic})\And \not\Omega_j(t,\Gamma^c_{vic}, \hat\Gamma^r_{vic}, \Gamma^{cr}_{vic}) \And \not\Omega_j(t,\Gamma^c_{vic}, \Gamma^r_{vic}, \hat\Gamma^{cr}_{vic})$}\label{alg2:cle_feasible}
            \State $\nabla_{c} = \nabla_{c} \cup \tau_j$ \Comment{Only \cleanup\ is feasible for $\tau_j$}
    
    \Else\label{alg2:not_feasible}
        \State $\nabla_{inf} = \nabla_{inf}\cup \tau_j$ \Comment{Security operations not feasible}
    \EndIf
    \EndFor

    \If{$|\nabla_{inf}| = \emptyset$}\label{alg2:if_feasible}
        \If{$|\nabla_{c}| = \emptyset \And |\nabla_{r}| = \emptyset \And |\nabla_{cr}| > 1$}\label{alg2:if_restart}
            \State $\Gamma^{cr}_{vic} = \hat\Gamma^{cr}_{vic}$ \Comment{Update the \restart\ queue}
        \ElsIf{$|\nabla_{c}| = \emptyset \And |\nabla_{r}| > 1$}\label{alg2:if_onlyrestart}
            \State $\Gamma^{r}_{vic} = \hat\Gamma^{r}_{vic}$ \Comment{Update the \onlyrestart\ queue}
        \ElsIf {$|\nabla_{c}| \neq \emptyset$}\label{alg2:if_clea}
        \State $\Gamma^c_{vic} = \hat\Gamma^c_{vic}$  \Comment{Update the \cleanup\ queue}
        \EndIf
        \Else\label{alg2:if_infeasible}
        \State Raise an alarm that security cannot be performed.
    \EndIf
\EndFor

\end{algorithmic}
\end{algorithm}

Algorithm~\ref{alg:restart_decision} leverages Theorem~\ref{theorem:delta_security} and Lemma~\ref{lemma:feasibility} to identify the optimal security operations that can be accommodated within the weakly-hard constraints of the control system. This algorithm is activated by anomaly detection mechanisms such as the RedZone principle discussed in Section~\ref{sec:cleanup_overhead}. When a security operation is initiated, the scheduler first determines the feasible set of security operations based on the priorities of the control tasks within the system. Each set of tasks utilized by Algorithm~\ref{alg:restart_decision} is organized in order of priority. Since we adhere to a priority-based approach for the victim task, the highest-priority tasks are minimally impacted by the security overhead.

The algorithm takes the set of victim tasks as an input, along with the set of all tasks in the system. Line~\ref{alg2:for_vic} loops over all the flagged victim tasks in the system. The loop iterates in order of higher to lower priority of the victim tasks. Lines~\ref{alg2:nabla_cle},~\ref{alg2:nabla_res}, and~\ref{alg2:nabla_inf} are temporary sets to store the set of system tasks that can accommodate an added security overhead. The set $\nabla_{inf}$ is the set of tasks for which an added security overhead will not be feasible. In Lines~\ref{alg2:res_set} and~\ref{alg2:if_restart}, the current victim task is added to the temporary queues for the security operations.

The loop in Line~\ref{alg2:for_all} iterates through all the other tasks in the system to check if $\Gamma^c_{vic}$, $\Gamma^r_{vic}$, or $\Gamma^{cr}_{vic}$ can be feasibly added to the system without affecting each of the other tasks in the system. For each of the tasks, the scheduler checks four conditions:\begin{enumerate}
    \item If a \restart\ is feasible;
    \item If a \onlyrestart\ is feasible;
    \item If a \cleanup\ is feasible;
    \item If neither of the added security overheads is feasible.
\end{enumerate}

Line~\ref{alg2:res_feasible} checks if the \restart\  is feasible and if the victim task is of higher priority than the current task being checked. Since a restarted lower-priority task will not interfere with the execution of the checked task, the~\onlyrestart\ overhead is only counted if the victim task is of higher priority. Line~\ref{alg2:onlyrestart_feasible} checks for the feasibility of ~\onlyrestart\ operations, and Line~\ref{alg2:cle_feasible} checks for the feasibility of adding only the \cleanup\ operation described in Algorithm~\ref{alg:cleanup_mechanism}. Line~\ref{alg2:not_feasible} adds the checked task to a set of tasks for which a security overhead is infeasible. Any security operation is only added to the system if Line~\ref{alg2:if_feasible} is true (i.e., $|\nabla_{inf}| = \emptyset$). If not, Line~\ref{alg2:if_infeasible} notifies the admin that the tasks have been marked as anomalous, but no security operation is feasible. If Line~\ref{alg2:if_restart} is true, then \restart\ is feasible, and $\Gamma^r_{vic} = \hat\Gamma^r_{vic}$. Similarly, $\Gamma^c_{vic} = \hat\Gamma^c_{vic}$ if Line~\ref{alg2:if_clea} is true.

\begin{figure*}[t]
    \begin{subfigure}[t]{0.3\textwidth}
\includegraphics[width=0.8\columnwidth]{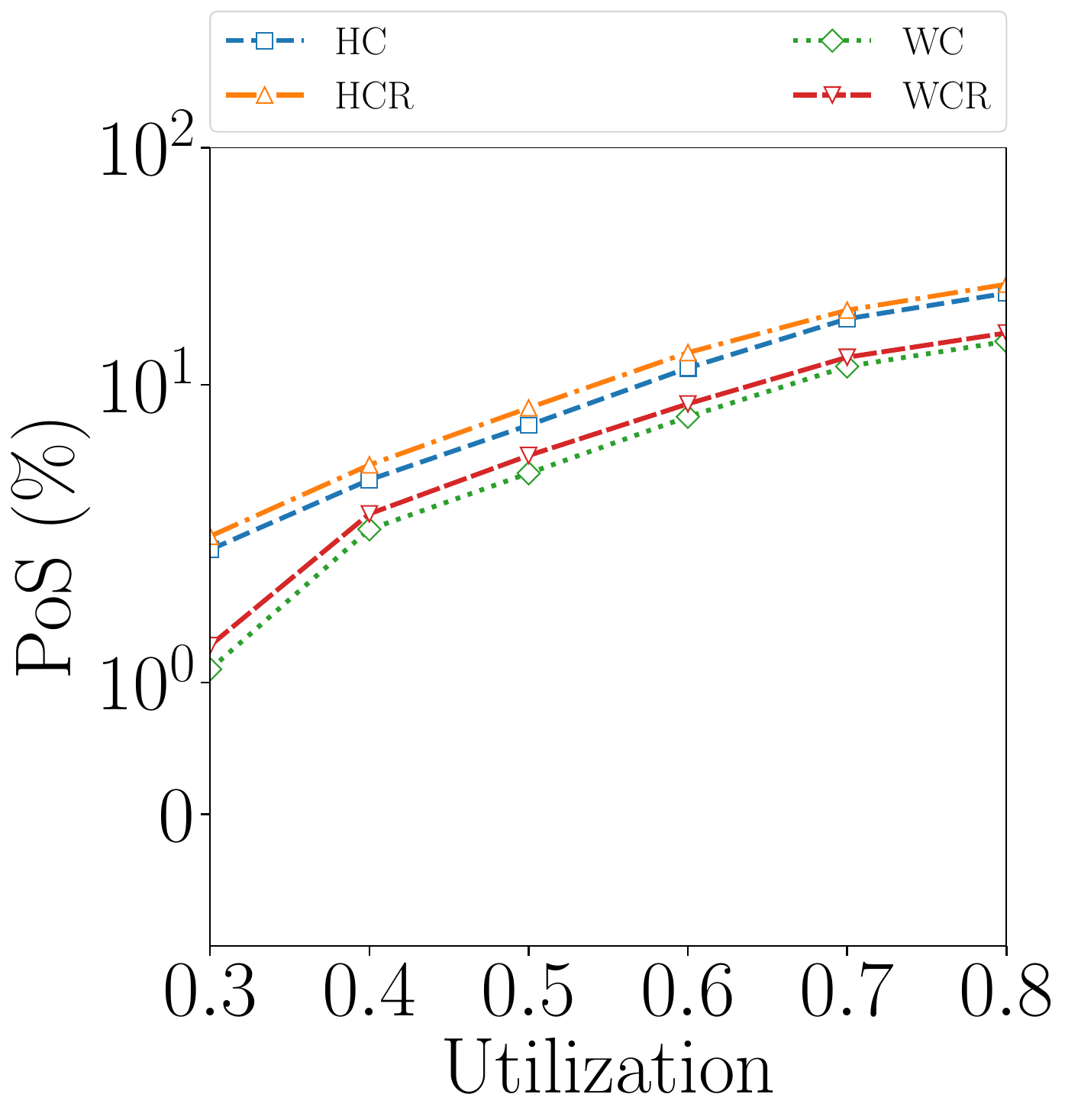}
\caption{Comparing Overheads}
\label{fig:feasibility_res_vs_clea}
    \end{subfigure}\hfill
     \begin{subfigure}[t]{0.3\textwidth}
\includegraphics[width=0.8\columnwidth]{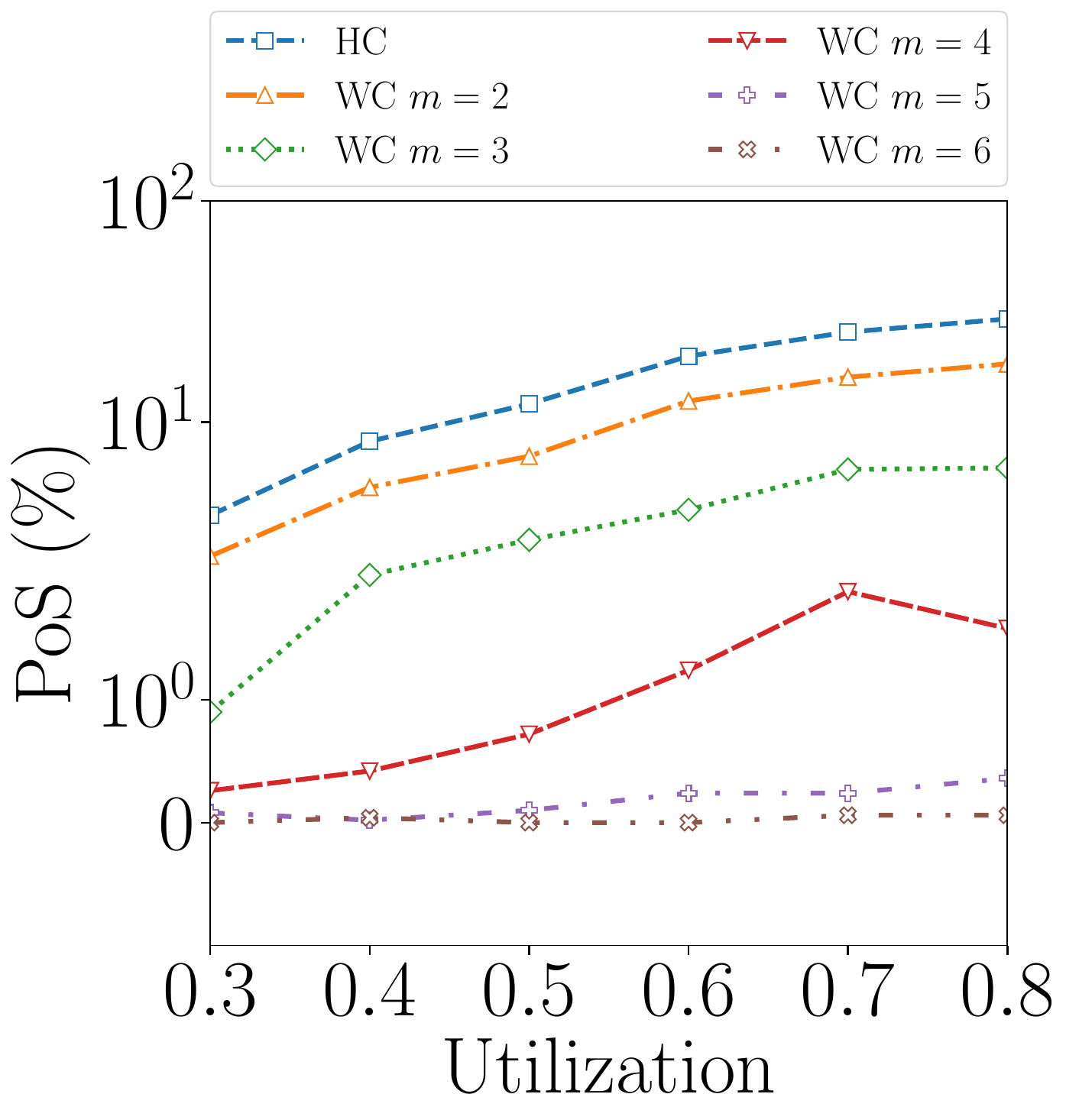}
\caption{Comparing Miss Tolerance}
    \label{fig:feasibility_vs_m}
    \end{subfigure}\hfill
      \begin{subfigure}[t]{0.3\textwidth}
     \includegraphics[width=0.8\columnwidth]{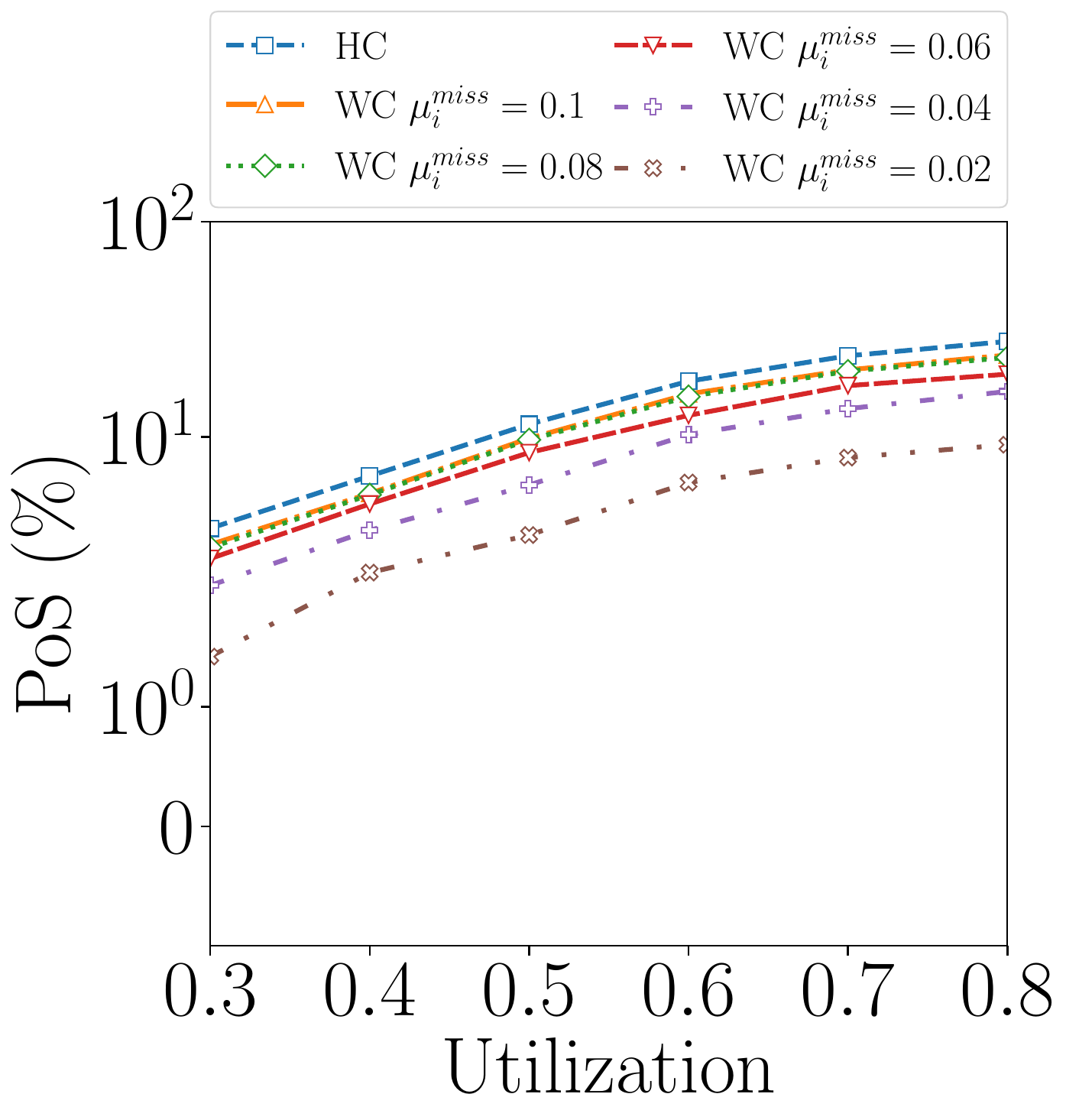}
        \caption{Comparing Miss Probability}
    \label{fig:feasibility_vs_util}
    \end{subfigure}
    \caption{Comparing the price of security (denoted by PoS; y-axis in the plots) at different levels of system constraints and security overheads. In all the plots, the y-axis is in $\log$ scale for better visualization. To reduce clutter in the plots, we have codified the configuration into three characters: \textbf{H:} Hard, \textbf{W:} Weakly-hard, \textbf{C:}~\cleanup, and \textbf{R}~\onlyrestart. So, \textbf{WCR} is a weakly-hard system with~\restart\ security operation. Figure \ca Compares the price of security in hard and weakly-hard systems for \cleanup\ and \restart\ operations. Figure \cb compares the price of added security in a weakly-hard system with that of a hard real-time system. With higher system utilization, the system incurs a higher cleanup cost. The weakly-hard constraint, $(m=2, k=20)$, is more feasible in terms of security compared to hard, \ie $(m=0, k=1)$ constraint systems. Finally, 
     figure \cc shows the price of security at different $(m,k)$. With a constant $\mu_i$ and a $5\%$ miss rate, task sets with $m > 4$ exhibit less than $2\%$ overhead even at $80\%$ system utilization. A hard constraint shows a sharp increase in the price of security with increasing utilization.
    }
\end{figure*}

\section{Evaluation}\label{sec:eval}

\subsection{Taskset Generation}

We evaluate the price of security using synthetic task sets. For the experiments presented in this section, we generated $5000$ task sets, each containing $10$ tasks. The tasks have unique priorities assigned according to the rate-monotonic order (shorter period implies higher priority). The periods $T_i$ are randomly assigned from $10$ ms to $100$ ms. Then we set the utilization levels at $30\%$ to $80\%$, and at each utilization level, we generate random weights for each task that determine the total portion of the utilization assigned to that task. We then calculate the value of $C_i$ by multiplying the target total utilization and the random weight for each task. To guarantee that the total utilization is consistent, we then multiply $C_i$ with $\frac{\text{Utilization Level}}{\sum \frac{C_i}{T_i}}$ to get the final value of $C_i$, which guarantees that the execution times are scaled to meet the total utilization level. 
For the set of tasks randomly generated, we pick task sets that are feasible based on response time analysis~\cite{audsley1993applying,lehoczky1990fixed}. The Performance threshold $J_i^{TH}$ is uniformly selected from $[100\ \text{ms}, 150\ \text{ms}]$. The baseline performance for all the following experiments is $100\%$, ensuring that the security overhead is compared across task sets that are otherwise $100\%$ feasible. The individual experiments list other related parameters.

\subsection{Results}

We study the price of security on the following two fronts: \ca hard vs weakly-hard requirements including the overheads of cleanup\ and \restart\ operations, and \cb miss probability and impact of priority assignments.

\subsubsection{Price of Security Under Hard vs. Weakly-Hard Requirements} \label{sec:eval_h_vs_wh}

The following experiments compare the price of security for different operations under hard and weakly-hard constraints. We also evaluate the price of security at different $(m,k)$ constraints.

\paragraph{Overhead of \cleanup\ and \restart}

In the first set of experiments, we compare the price of the \cleanup, which adds the lowest security overhead, and \restart\ which adds the highest overhead, under hard and weakly-hard constraints. The results of this experiment are shown in Figure~\ref{fig:feasibility_res_vs_clea}. We observe that weakly-hard constraints outperform the hard constraints considered in prior work on scheduler-based security. For this experiment, we fixed the value of $\psi_i = 0.2$. We notice similar trends in Figure~\ref{fig:feasibility_res_vs_clea} and Figure~\ref{fig:feasibility_vs_util}, where the security overhead in a weakly-hard system is consistently lower than that of a system with hard constraints.

\paragraph{Miss Tolerance}

In Figure~\ref{fig:feasibility_res_vs_clea}, the $\mu_i$ patterns are generated by randomly sampling from a set of binary numbers with $5\%$ miss rate. For a system with more frequent deadline misses, \repose\ still shows a substantially lower cost than hard constraints. The performance can be further improved by adjusting system design parameters, such as the $(m,k)$ constraints and the frequency of deadline misses during normal system operation.

Figure~\ref{fig:feasibility_vs_m} shows a comparison of the price of security in systems with hard constraints to that of weakly-hard constraints at different levels of \emph{miss tolerance}, i.e., at different $(m,k)$ ratios. Based on the observation in Figure~\ref{fig:feasibility_res_vs_clea} and Figure~\ref{fig:feasibility_vs_m}, the system designer can adjust design time parameters such as the $(m,k)$ constraint of the system using prior work exploring the safety of control systems under deadline misses~\cite{vreman2021stability}.

In Figure~\ref{fig:feasibility_vs_m}, we also observe that the price of security has a steeper positive slope with lower values of $m$, showing that higher system utilization has a more substantial impact on the security of systems with tighter constraints. With higher miss tolerance, security operations can be scheduled more frequently. Note that a security event is comparatively infrequent, and even tighter bounds can yield better average performance in some cases. However, in these evaluations, we measure the worst-case timing performance to demonstrate the upper bounds on the price of security.

Figure~\ref{fig:feasibility_vs_util} compares the price of security of task models with different levels of deadline meet and miss ratios. In this comparison, we assume that only one task is flagged as a victim for a randomly selected value of $t$. We assume all required security operations are being executed rather than selectively executing only the feasible operations. The price of security trend provides an insight into what fraction of tasks can tolerate added security overhead. In this experiment, we consider the overhead from \cleanup. Hence, the overhead added to the task execution would be $\psi_iC_i$, where $\psi_i$ is set to $20\%$. Since, the hard constraint is a special case of weakly-hard system, in the following experiment, we set the value $m=0$ in $(m,k)$ constraint of the task to get the overhead in a hard real-time system. 

For the experiment plotted in Figure~\ref{fig:feasibility_vs_util}, we set the $(m,k)$ bounds of each task constant at $m=2; k=20$, then set the value of $\mu_i$ as a binary string of length $20$ randomly sampled from a set of $1000$ binary numbers. For each of the weakly-hard models, we randomly sample the set of binary numbers by setting the total number of
$0$'s between $2\%$\ to $10\%$. We define $\mu_i^{miss}$ as the miss probability, which we define as the probability of miss, or the percentage of $0$'s in an arbitrary $\mu_i$ string.
This experiment demonstrates how $\mu_i^{miss}$ in a weakly-hard system affects security overhead. With higher miss probability, the weakly-hard constraint shown in Equation~\ref{eq:omega_define} will face tighter bounds due to a higher frequency of deadline misses (and hence,  show a higher price of security).

Note that the hard constraint in Figure~\ref{fig:feasibility_vs_util} still uses online security, and the \cleanup\ is triggered when the task goes into a RedZone. In contrast, for a task set with a security mechanism added to all tasks, the total performance cost of the system would substantially increase, and the total overhead would have a significantly higher price of security than weakly-hard systems. Algorithm~\ref{alg:restart_decision} selects an optimal schedule for security operations to minimize the price of security and maximize the security of the system, even on a higher utilization system. For task sets where the price of security is high enough to make all security operations infeasible, \repose\ raises an alarm and notifies the admin that the system is in an unsecured state.

\begin{figure*}[t]
    \begin{subfigure}[t]{0.3\textwidth}
    \includegraphics[width=\columnwidth]{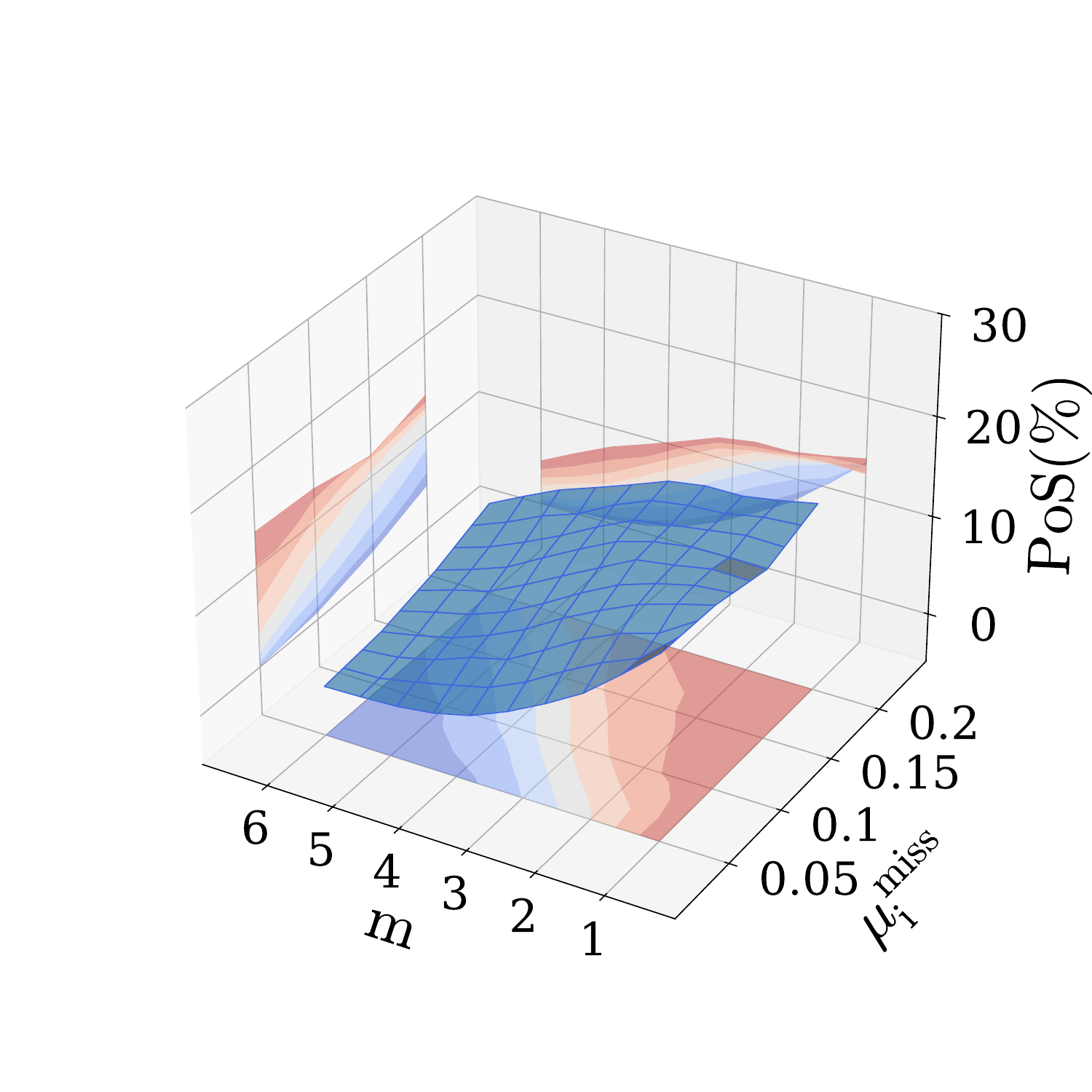}
        \caption{Impact of miss tolerance and miss probability on PoS with rate-monotonic priority}
\label{fig:missfreq_vs_m_vs_feasibility}
    \end{subfigure}
    \hfill
    \begin{subfigure}[t]{0.3\textwidth}
\includegraphics[width=\columnwidth]{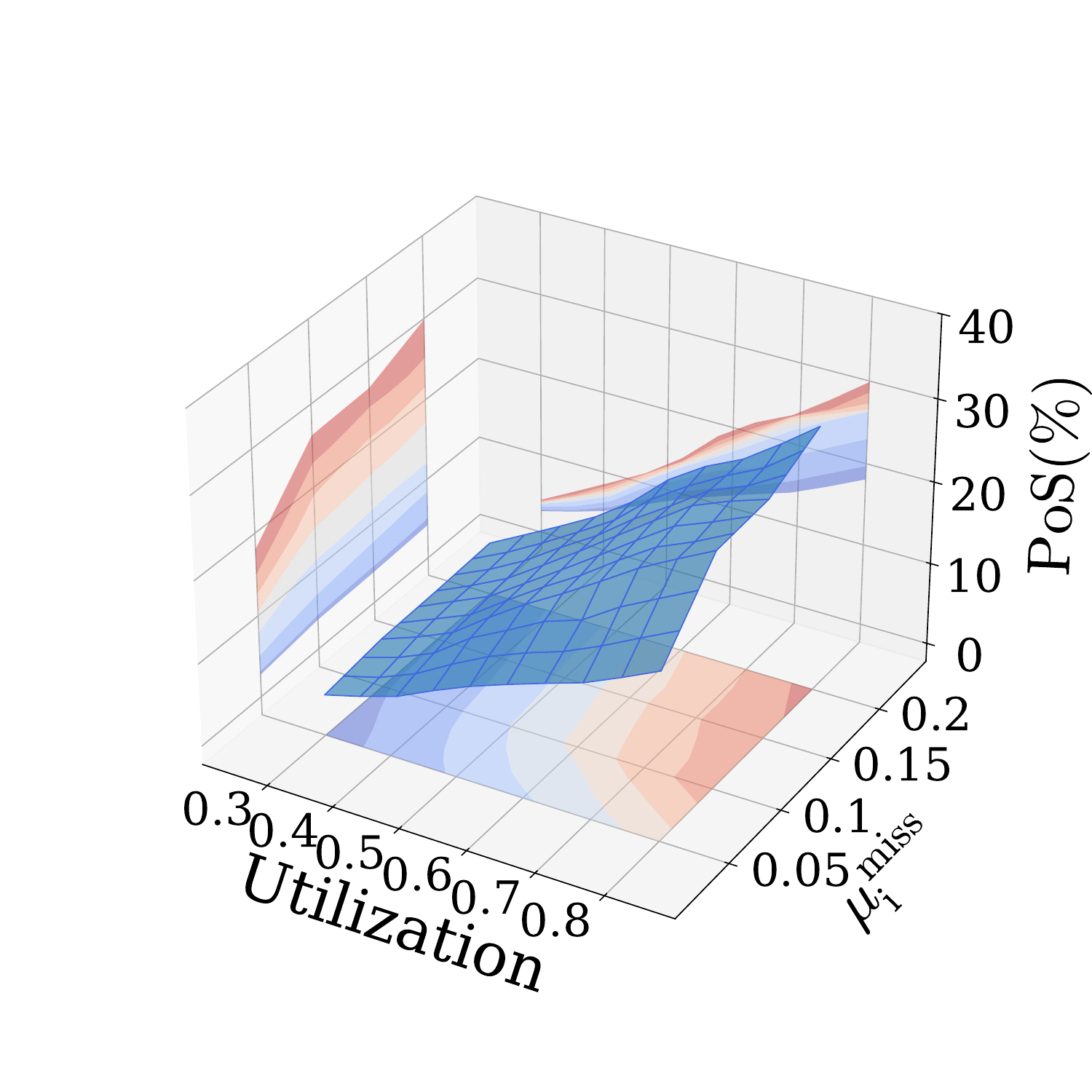}
    \caption{Impact of miss probability and utilization on PoS with rate-monotonic priority}
    \label{fig:feasibility_rms}
    \end{subfigure}
    \hfill
    \begin{subfigure}[t]{0.3\textwidth}\includegraphics[width=\columnwidth]{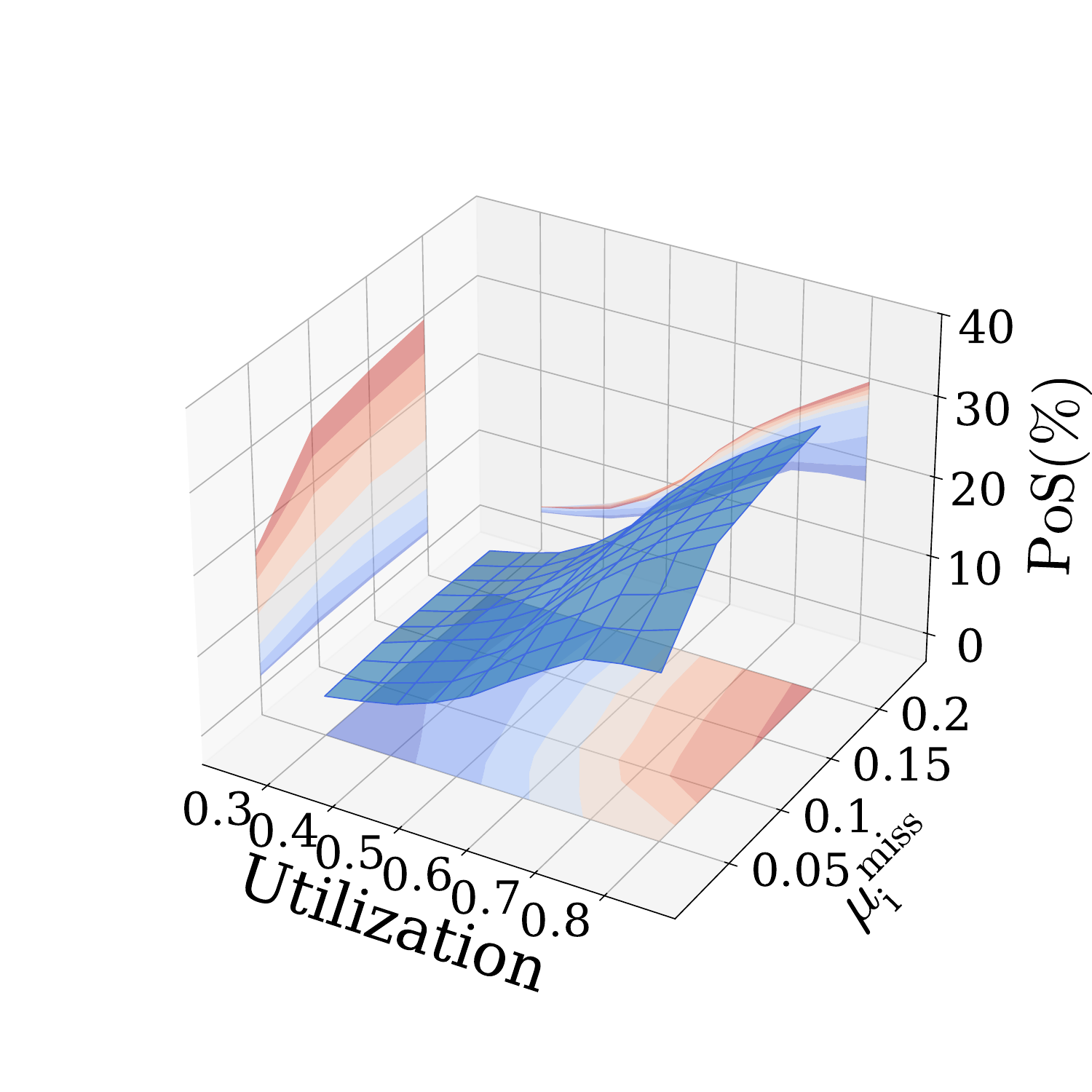}
    \caption{Impact of miss probability and utilization with arbitrary fixed-priority.}
    \label{fig:feasibility_afpp}
    \end{subfigure}
    \caption{Comparing price of security (PoS) with different scheduling policies, miss tolerance ($m$), and miss probability($\mu_i^{miss}$). The plots illustrate the following evaluations: \ca PoS as a function of the miss probability and the miss tolerance with rate-monotonic priority---miss tolerance of the system has a higher impact on security feasibility;
    \cb PoS as a function of $\mu_i^{miss}$ and utilization with rate-monotonic priority; and 
    \cc PoS as a function of $\mu_i^{miss}$ and utilization with arbitrary fixed-priority.}
\end{figure*}

\subsubsection{Miss Probability with Varying Weakly-Hard Requirement and Impact of Priority Assignment}

Figure~\ref{fig:missfreq_vs_m_vs_feasibility} shows a comparison of the price of security from changing miss tolerance and miss probability. For this experiment, the utilization of the system was fixed at $50\%$, and the priority assignment was rate-monotonic~\cite{lehoczky1989rate}. We observe that the miss tolerance has a higher impact on the cost of security than the miss probability. This result can also be attributed to the fact that the miss tolerance rate is higher than the miss probability rate. In a given $\mu_i$ of length $k$, if $\frac{m_i}{k_i} > \mu_i^{miss}$, then the system shows substantially higher feasibility of security overhead compared to other scenarios.

Based on the observation in Figure~\ref{fig:missfreq_vs_m_vs_feasibility}, a system designer can extend the $(m,k)$ bounds of the system to accommodate better security in cases where the probability of a deadline miss is much higher or in places where system overload is more frequent. Prior work on developing control systems with deadline miss tolerance~\cite{vreman_stability_2021,pazzaglia_dmac_2019} can be combined with \repose\ to enhance the system's security posture without compromising performance.

\begin{figure*}
    \centering
    \includegraphics[width=0.7\linewidth]{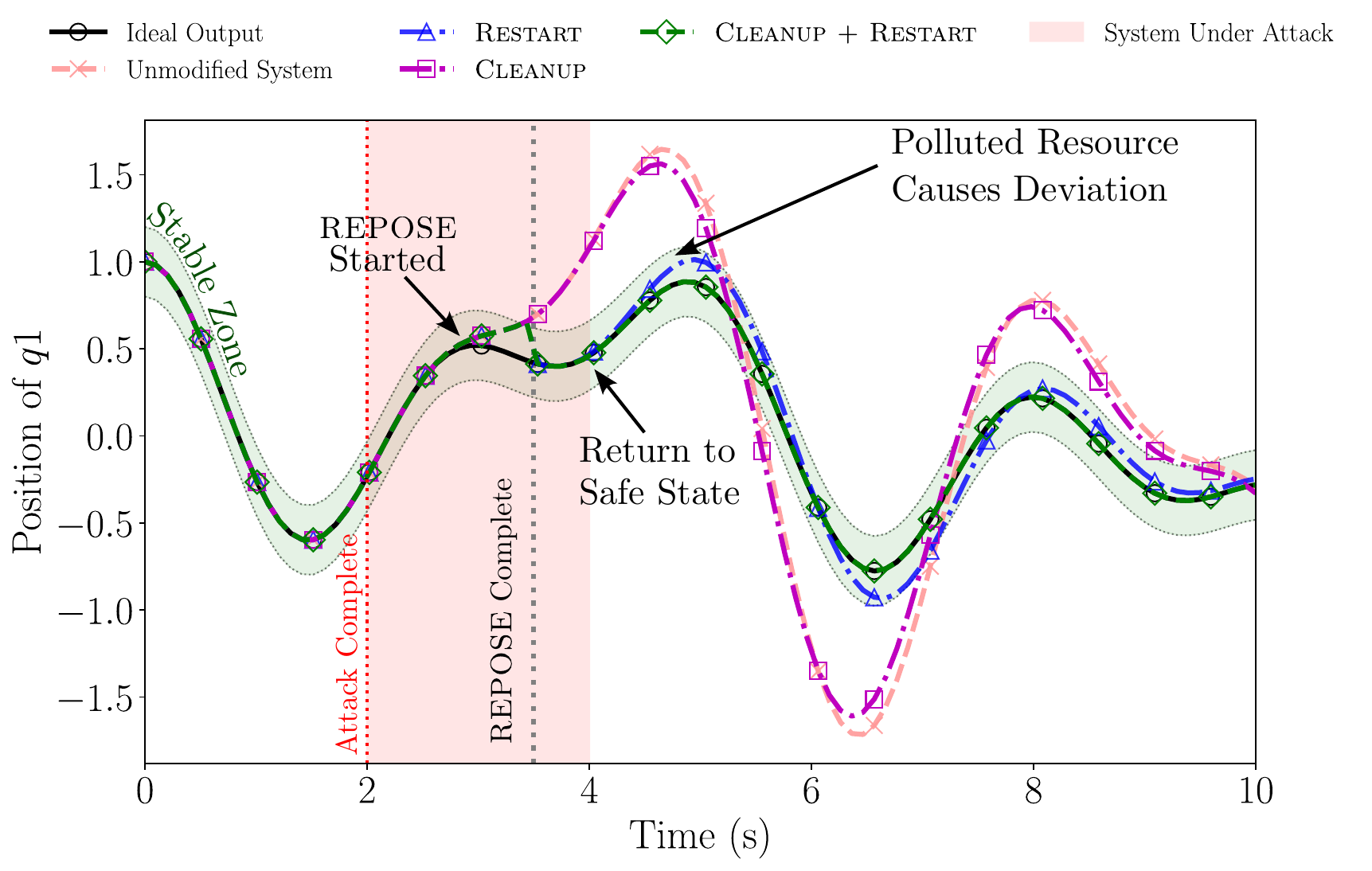}
    \caption{Different security operations of~\repose\ to recover the system into a safe state. The system was under attack at time $t = 2 $~s. Using an IDS~\repose\ triggered security operations at $t = 3$~s. In this experiment, the time overhead for \repose is $0.5$~s. After the security operations are performed, the system returns to a secure state.}
    \label{fig:ctrl_experiment}
\end{figure*}

In Figure~\ref{fig:feasibility_rms} and Figure~\ref{fig:feasibility_afpp}, we show the security overhead on weakly-hard systems with a miss tolerance $m = 2$ using rate-monotonic and arbitrary fixed-priority~\cite{lehoczky1990fixed} scheduling policies, respectively. We observe that the scheduling policies have a minimal impact on the price of security. We also notice that the arbitrary priority assignment algorithm performs slightly better than the rate-monotonic scheduling policy. This observation can be attributed to the difference in how priority is assigned to tasks. In rate-monotonic, tasks with a lower period are assigned a higher priority. Due to the higher frequencies, these tasks also have lower $J^{TH}$ values; hence, the addition of any security cleanup has a greater impact on the higher-priority tasks in rate-monotonic scheduling. In arbitrary priority assignments, on the other hand, tasks have different priority assignments independent of their periods, making them more resilient to added security.

The plots in Figures~\ref{fig:feasibility_rms} and~\ref{fig:feasibility_afpp} illustrate the frequency at which \repose\ will skip security scheduling and notify the administrator of a potential attack, or choose to trade off performance to accommodate the security operations. Prior work on scheduling-based attacks supports this finding, as predictable timing makes the system more vulnerable and prone to side-channel attacks~\cite{nasri2019pitfalls}. Weakly-hard constraints still minimize the impact of predictable rate-monotonic scheduling, as the security and restart overheads affect lower-priority tasks more than higher-priority tasks due to the priority-based scheduling of security operations, as shown in Algorithm~\ref{alg:restart_decision}. 

We also notice that the miss probability has a less significant impact on the price of security with a constant value of $m$, demonstrating the effectiveness of having the $(m,k)$ bounds. At low utilization, the miss probability has minimal impact, as security can still be scheduled even with hard constraints. At higher utilization, the system needs to use the weakly-hard constraints more frequently, resulting in a higher impact of the miss probability. Hence, even in systems that frequently face overloads, security can be improved by bounding the miss tolerance $m$. In statically added security mechanisms~\cite{standaert2025limited}, these task sets would not be feasible, resulting in a much more pessimistic schedule that impacts control performance.

We evaluate the price of security under weakly-hard and hard real-time constraints using Theorem~\ref{theorem:delta_security} and Lemma~\ref {lemma:feasibility}. Note that this performance outcome study examines whether security operations would be scheduled by~\repose, rather than the system's schedulability itself. If the price of security is infeasible,~\repose\ will notify the administration that the security is in an unsecured state, as shown in Algorithm~\ref{alg:restart_decision}.

\section{Case Study: \repose\ Operation in a Real-Time Control System}

\begin{figure}[!h]
    \centering
    \includegraphics[width=0.7\linewidth]{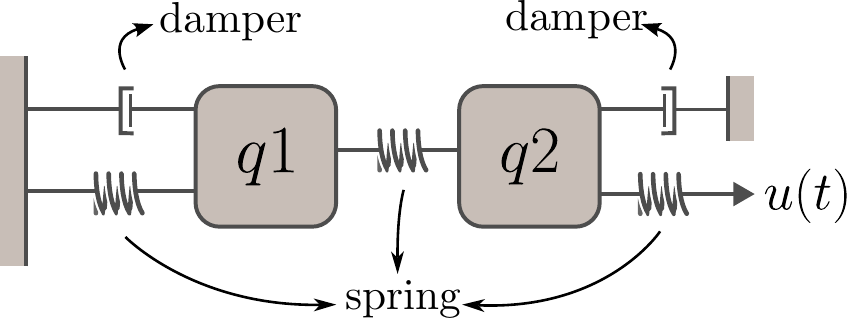}
    \caption{Two-Mass Spring-Damper system with a $u(t)$ input.}
    \label{fig:spring_damper}
\end{figure}

We also present a case study of \repose\ in a control system that is representative of the plant model described by Equation~\ref{eq:state_rate} and Equation~\ref{eq:control_output}. We study the price of security under different security operations in a two-mass spring-damper experiment shown in Figure~\ref{fig:spring_damper}. In the given figure, we assume the $u(t)$ to be a sinusoidal input, which is a common assumption in the evaluation of control systems. When an input is applied to the system, the position of $q_1$ and $q_2$, \ie $y(t)$, depends on the rate of change of state $\dot x$ shown in Equation~\ref{eq:state_rate} and the control input $u(t)$. Using the classic control system model allows us to demonstrate the control system-based mapping of security in RT-CPS that is utilized by~\repose. We simulated the system behavior using the \texttt{python-control}\cite{python-control2021} library and added modifications to $u(t)$ at a randomly selected time to simulate a resource-based attack and observe the system behavior under different~\repose\ security operations.

Here, we note that the control input $u(t)$ is similar to a buffered input used by RT-CPS. This control input is equivalent to a sensor input used by safety-critical systems to make real-time decisions. For ease of demonstration, we focus on the position of $q_1$ only, as $q_2$ always shows a complementary position and can be calculated from the state of $q_1$. At any given time $t$, we assume that the current position of the system only depends on the control input and the state of the system, and the position is calculated at the time of reading the input.

Let us assume that an attacker pollutes the input either through a side-channel or a data injection attack in the cache at time $t=2$~s. The state of the task changes at time $t=3$~s, as the task can only process the input in the next cycle of execution. We note that the system deviates from its ideal expected execution due to the attack. The IDS detects this deviation and triggers \repose. The plots in Figure~\ref{fig:ctrl_experiment} show the price of security for different approaches that can be taken by~\repose\ in this situation.

The~\cleanup\ operation here is to clear the $u(t)$ buffer and use the current fresh input of the system so that the impact of the attack is mitigated. The~\onlyrestart\ operation discards the old state of the system and re-instantiates the task, so the current state is calculated based on the fresh input. We observe that an attack can potentially put the system in an unsafe state. However, with the ~\restart\ and ~\onlyrestart\ approaches, we were able to bring the system back to a safe execution state (see the position at around $3.5$~s). However, notice that the state of the system with~\onlyrestart\ eventually deviates slightly from the baseline ideal output because even with a clean system state, the output is still impacted by the cache pollution caused by the attack. Depending on the tolerance of the system, this slight deviation might or might not put the system in an unsafe state. In this experiment, we assume that the slight deviation is within the tolerable bounds and hence the system can function normally. We also notice that the combined ~\restart\ perfectly merges back into the ideal output curve as the fresh state of the system and a fresh, cleaned-up buffer completely removes the impact of a cache pollution attack. 

We observe in Figure~\ref{fig:ctrl_experiment} that the curves corresponding to an unmodified system (dashed light pink) and the ~\cleanup\ approach (dot-dashed purple) put the system in a prolonged, unsafe state. This observation is due to the incorrect state of the system that has already been calculated, and the system has already been put in an unsafe state. Without re-instantiating the state of the system, the system will stay in an unsafe state even if it slightly deviates from a completely unsecured system.

The control system experiment provides a depiction of the potential security impact on the system. Due to the weakly-hard real-time nature, the system can only tolerate bounded deadline misses, making it crucial for~\repose\ to take into account the available time and the security level it can achieve within that time. We observe from the experiment that the highest security is provided by~\restart\ mechanism (dashed green), but the system can still achieve tolerable performance deviation with a ~\onlyrestart\ system (dot-dashed blue). The lowest price of security is from the~\cleanup\ operation, but the security impact (coarse-grained cleaning) might make the system unsafe.

\section{Discussion}\label{sec:discussion}

\cpar{Reactive and Proactive Security.}
\repose\ uses a reactive security mechanism to execute security cleanup only when a security event is detected. However, with the rise of cyber attacks against RT-CPS~\cite{hasan2024sok}, numerous evasion techniques can be employed against existing detection mechanisms~\cite{banerjee2025timing,banerjee2025evading}. To further enhance the security of RT-CPS, \repose\ can be used in reactive and proactive security triggers. Prior work has extensively explored different mechanisms to add proactive security in RT-CPS~\cite{hasan_contego_2017,hasan2022beyond,abdi2018guaranteed,banerjee2022secure}.

The theoretical framework used by \repose\ to measure feasibility can be combined with the offline security parameter calculation to execute Algorithm~\ref{alg:restart_decision} at online detection, as well as at pre-calculated points in the timeline, to enable the cleanup of tasks that evade detection mechanisms.
\repose\ can also be triggered at random times to improve security. Since \repose\ always checks for the feasibility of security overhead before scheduling, triggering \repose\ at random times would not impact the overall schedulability of the system. Future work can use a randomized or pre-calculated execution of \repose.

\cpar{Implementing \repose\ in RTOS.}
\repose\ can be implemented using common scheduler functions provided by most RTOSes. \repose\ mainly requires the scheduler to be able to know how much of a task's budget it has already used, which is common in most RTOSes, including RTEMS~\cite{bloom2020real} and FreeRTOS~\cite{barry2008freertos}. The APIs required for implementation, as discussed in Section~\ref{sec:cleanup_mech}, are commonly available.  Since the requirements are only at the software level, \repose\ can be implemented on COTS devices without modification, using most open-source RTOS implementations. Our future work will explore integrating \repose\ into a control system environment running RTEMS and EPICS~\cite{lee2016distributed}, and study the cost of security in a production-level control system environment.

\cpar{Security Cleanup Mechanisms.}
The security cleanup shown in this paper is a generic cache cleanup mechanism that can be implemented using existing system calls in popular RTOSes. Different security mechanisms would have different timing and space impacts on the overall execution. We synthetically explore the impact of security by considering security overhead as a percentage of the task's execution time. One critical aspect is to examine the range of security mechanisms that can be integrated with \repose\ and to identify how their associated overheads can be minimized to strengthen the security posture of RT-CPS. We intend to explore this aspect in future work.

There exist other system resources that are allocated and used by real-time control systems. Shared I/O buffers are examples of such resources. In resources used by multiple tasks simultaneously, the cleanup mechanism of one task can affect others. \repose\ assumes that each security operation only impacts the associated task. However, because the cleanup operation affects all tasks in the system, the cleanup executed by \repose\ can be integrated with systems that employ multiple shared resources.

\section{Related Work}

The work in the paper is related to two broad research areas: \ca scheduler-based security and \cb maintaining physical safety in the presence of a deadline miss. 

\subsection{Scheduler-based Security} Security-focused task scheduling has been approached in two categories: randomization-based scheduling and deterministic scheduling overhead. Randomized scheduling~\cite{yoon2016taskshuffler, kruger2018vulnerability} is generally targeted at mitigating side-channel attacks by adjusting the execution times of real-time tasks in the system. However, due to the timing constraints of real-time systems, randomization-based scheduling has been proven ineffective against persistent attacks~\cite{nasri2019pitfalls}. In contrast, deterministic scheduling overhead mechanisms, such as SchedGuard~\cite{chen2021schedguard, chen2023schedguard++}, or restart-based system recovery~\cite{banerjee2022secure,abdi2018guaranteed,hounsinou2021work} add security overhead to each task in the system to perform security cleanup, thereby mitigating the impact of attacks. In these mechanisms, the scheduling overhead can substantially impact the performance due to a significant increase in system utilization.

Researchers explore integrating security as independent periodic tasks~\cite{hasan2016exploring,hasan_contego_2017}. To improve the performance of task models with security mechanisms, prior work has introduced security-focused scheduling~\cite{baruah2022security,raadia2024improved,baruah2023scheduling}, which propose scheduler models to efficiently schedule tasks with integrated security mechanisms. However, these techniques impose a constant security overhead on each task, resulting in a significantly higher security cost. Besides, they do not consider control aspects. Contego-C~\cite{hasan2022beyond} provides a mechanism for integrating security into RT-CPS using design-time parameters. None of these works considers weakly-hard requirements, which makes security costs less feasible in highly utilized systems.

\subsection{Maintaining Physical Safety Under Deadline Miss}

The scheduler-based security approach assumes that all tasks in the system must meet their deadlines; therefore, the system must be a hard real-time system to ensure its safety. Prior work has demonstrated that, in practical systems, tasks can often miss a deadline without significantly affecting system safety. There has been rigorous research on handling these deadline misses without compromising safety~\cite{salamun_weakly_2023}. Weakly-hard system models, first introduced in~\cite{bernat2001weakly}, provide a method for bounding the number of tolerable deadline misses within a given time window. The concept of weakly-hard tasks has been widely used in control systems to understand and design control systems that can tolerate bounded deadline misses~\cite{pazzaglia2018beyond, vreman2021stability}.

Security enhancements in RT-CPS can also introduce sufficient overhead and may cause a task to miss its deadline. Prior work has shown that the system safety can still be analyzed and maintained using the physical safety properties. For instance, the security overhead from restart-based recovery can be tolerated by RT-CPS if it is bounded by the system's physical properties, such as inertia~\cite{abdi2018guaranteed, abdi2018preserving}.

In this work, we integrate security into a weakly-hard control system and provide a theoretical framework assessing the feasibility of security integration at any arbitrary time $t$ during the system's execution. In contrast to prior work, \repose\ does not require adding a security mechanism to each task or pre-calculation of checkpoints to execute security cleanup mechanisms proactively. Instead, \repose\ is more \textit{adaptive} and \ca triggers a security check only when anomalous behavior is detected in a task, and \cb uses a priority-based scheduling of the weakly-hard task model to enable security operations in higher-priority tasks without compromising performance.

\section{Conclusion}\label{sec:conclusion}

In this paper, we introduce an online security mechanism, \repose\ that integrates security operations in weakly-hard real-time control systems. This paper presents analytical bounds that can be utilized for integrating online security mechanisms into control systems, leveraging bounds on the control delay parameter. \repose\ can be implemented in COTS components using APIs that are commonly available in most RTOS. Through an evaluation of a synthetic task set, we demonstrate the impact of different system parameters on the feasibility of security. The evaluations reveal a clear trade-off between security and the system's feasibility, which can be dynamically handled by \repose\ and notified to the administrator if security operations are infeasible. We believe \repose\ will provide valuable insights that further the study of the safety-security tradeoff in safety-critical RT-CPS.

\bibliographystyle{plainurl}
\bibliography{ref.bib}
\end{document}